\documentclass[prd,amsmath,amssymb]{revtex4}
\def\be{\begin{equation}}
\def\ee{\end{equation}}

\usepackage{graphicx}
\usepackage{natbib}
\usepackage{epsfig}
\usepackage{amsmath}
\begin{document}
\title{Astrometric Effects of a Stochastic Gravitational Wave Background}
%

\author{Laura G.\ Book}
\affiliation{\itshape Mail Code 350-17, California Institute of Technology, Pasadena, California 91125, USA}
\author{\'{E}anna \'{E}. Flanagan}
\affiliation{\itshape Center for Radiophysics and Space Research, Cornell University, Ithaca, New York 14853, USA and Newman Laboratory for Elementary Particle Physics, Cornell University, Ithaca, New York 14853, USA}

%
%
\begin{abstract}

A stochastic gravitational wave background causes the apparent positions of distant sources to fluctuate, with angular deflections of order the characteristic strain amplitude of the gravitational waves. These fluctuations may be detectable with high precision astrometry, as first suggested by Braginsky et al. in 1990.  Several researchers have made order of magnitude estimates of the upper limits obtainable on the gravitational wave spectrum $\Omega_{\rm gw}(f)$, at frequencies of order $f\sim 1 \, {\rm yr}^{-1}$, both for the future space-based optical interferometry missions GAIA and SIM, and for very long baseline interferometry in radio wavelengths with the SKA. For GAIA, tracking $N \sim 10^6$ quasars over a time of $T \sim 1 $ yr with an angular accuracy of $\Delta \theta \sim 10 \mu {\rm as}$ would yield a sensitivity level of $\Omega_{\rm gw} \sim (\Delta \theta)^2/(N T^2 H_0^2) \sim 10^{-6}$, which would be comparable with pulsar timing. 

In this paper we take a first step toward firming up these estimates by computing in detail the statistical properties of the angular deflections caused by a stochastic background. We compute analytically the two-point correlation function of the deflections on the sphere, and the spectrum as a function of frequency and angular scale.  The fluctuations are concentrated at low frequencies (for a scale invariant stochastic background), and at large angular scales, starting with the quadrupole. The magnetic-type and electric-type pieces of the fluctuations have equal amounts of power.

\end{abstract}
\maketitle
\pagestyle{empty}
%
%
\section{Introduction and Summary of Results}  \label{sec:intro}

\subsection{The Stochastic Gravitational Wave Background}

There is great interest in detecting or constraining the strength of stochastic gravitational waves (GWs) that may have been produced by a variety of processes in the early Universe, including inflation. The strength of the waves is parameterized by their energy density per unit logarithmic frequency divided by the critical energy density, $\Omega_{\rm gw}(f)$.  Current observational upper limits include (i) the constraint $\Omega_{\rm gw} \alt 10^{-13} (f / 10^{-16} \, {\rm Hz})^{-2}$ for $10^{-17}~{\rm Hz}~\alt~f \alt 10^{-16} \, {\rm Hz}$ from large angular scale fluctuations in the cosmic microwave background temperature \cite{Buonanno07}; (ii) the cosmological nucleosynthesis and cosmic microwave background constraint $\int d\ln f \: \Omega_{\rm gw}(f) \alt 10^{-5}$, where the integral is over frequencies $f \agt 10^{-15}$ Hz \cite{Smith06}; (iii) the pulsar timing limit $\Omega_{\rm gw} \alt 10^{-8}$ at $10^{-9} \, {\rm Hz} \alt f \alt 10^{-8} \, {\rm Hz}$ \cite{Jenet06}; (iv) the current LIGO/VIRGO upper limit $\Omega_{\rm gw} \alt 7 \times 10^{-6}$ at $f \sim 100 \, {\rm Hz}$ \cite{Abbott09}; and (v) the limit $\int d\ln f \: \Omega_{\rm gw} \alt 10^{-1}$ for $10^{-17} \, {\rm Hz} \alt f \alt 10^{-9} \, {\rm Hz}$ from very long baseline interferometry (VLBI) radio astrometry of quasars.

Many new techniques also promise future measurements of these primordial GWs. Firstly, it has been shown that such a GW background would leave a detectable signature in the polarization of the cosmic microwave background (CMB) \cite{Kamionkowski97,Seljak97}, which will be measured by many current and future observational efforts \cite{ACTPol10,Bicep10,CAPMAP08,CBI05,Clover08,CMBPol08,Ebex04,QUaD09,PIPER10,Polarbear10}. The planned space-based interferometer LISA will also set limits on the primordial stochastic gravitational wave background (SGWB) \cite{Kudoh06}. The planned successor to LISA, the Big Bang Observer, is a space-based interferometer mission designed primarily to detect the primordial SGWB \cite{Phinney03}. Finally, Seto and Cooray have suggested that measurements of the anisotropy of time variations of redshifts of distant sources could provide constraints of order $\Omega_{\rm gw} \alt 10^{-5}$ at $f \sim 10^{-12}$ Hz \cite{Seto06}. For more details on GWs, the search for them, and the SGWB, see the review articles \cite{Buonanno07, Allen97, Maggiore00}.

\subsection{High Precision Astrometry}

The possibility of using high precision astrometry to detect GWs has been considered by many authors \cite{gaia00, Braginsky90, Damour98, Fakir94, Flanagan93, Gwinn97, Jaffe04, Kaiser97, Linder86, Makarov10, Mignard10, Pshirkov08, Pyne96, Schutz09}. There was an early suggestion by Fakir \cite{Fakir94} that GW bursts from localized sources could be detectable by the angular deflection $\Delta \theta$ to light rays that they would produce. Fakir claimed that $\Delta \theta \propto 1/b$, where $b$ is the impact parameter. This claim was shown later to be false, and in fact the deflection scales as $1/b^3$ \cite{Kopeikin98, Damour98, 1999PhRvD..60l4002K, 2006CQGra..23.4299K}. Therefore the prospects for using astrometry to detect waves from localized sources are not promising \cite{Schutz09}.

However, the situation is different for a SGWB, as first discussed by Braginsky et al. \cite{Braginsky90}. For a light ray propagating through a SGWB, one might expect the direction of the ray to undergo a random walk, with the deflection angle growing as the square root of distance. However, this is not the case; the deflection angle is always of order the strain amplitude $h_{\rm rms}$ of the GWs, and does not grow with distance\footnotemark[1] \cite{Braginsky90, Kaiser97, Linder86}. \footnotetext[1]{It is sometimes claimed in the literature that the deflection angle depends only on the GWs near the source and observer. In fact, this is not true, as we discuss in the Appendix. A similar claim about the frequency shift that is the target of pulsar timing searches for GWs is also false in general.} Specifically, a SGWB will cause apparent angular deflections which are correlated over the sky and which vary randomly with time, with a rms deflection $\delta_{\rm rms}(f)$ per unit logarithmic frequency interval of [see Eq. \ref{eqn:2ptnn} below]

\be \delta_{\rm rms}(f) \sim h_{\rm rms}(f) \sim \frac{H_0}{f} \sqrt{\Omega_{\rm gw}(f)}. \label{eqn:deltrms} \ee

Suppose now that we monitor the position of N sources in the sky, with an angular accuracy of $\Delta \theta$, over a time T. For a single source, one could detect an angular velocity (proper motion) of order $\sim\Delta \theta/T$, and for N sources, a correlated angular velocity of order $\sim\Delta \theta/(T \sqrt{N})$ should be detectable. The rms angular velocity from (\ref{eqn:deltrms}) is $\omega_{\rm rms}(f)~\sim~f~\delta_{\rm rms}(f)~\sim~H_0~\sqrt{\Omega_{\rm gw}(f)}$, and it follows that one should obtain an upper limit on $\Omega_{\rm gw}$ of order \cite{Pyne96}

\be \Omega_{\rm gw}(f) \alt \frac{\Delta \theta^2}{N T^2 H_0^2}. \label{eqn:omapprox}\ee

\noindent This bound will apply at a frequency of order $f\sim1/T$. It will also apply at lower frequencies \cite{Pyne96} since the angular velocity fluctuations are white (equal contributions from all frequency scales), assuming a flat GW spectrum $\Omega_{\rm gw}~=$~const. The quantity that will be constrained by observations is roughly this total $\Omega_{\rm gw}$, $\int_{f \alt T^{-1}} d \ln f \Omega_{\rm gw}(f)$.

The advent of microarcsecond astrometry has started to make the prospects for constraining GW backgrounds more interesting. The future astrometry mission GAIA (Global Astrometric Interferometer for Astrophysics) is expected to measure positions, parallaxes and annual proper motions to better than $20 \: \mu\text{as}$ for more than $50\times 10^6$ stars brighter than $V \sim 16$ mag and 500 000 quasars brighter than $V \sim 20$ mag \cite{gaia00}. Similarly the Space Interferometry Mission (SIM) is expected to achieve angular accuracies of order $10 \: \mu\text{as}$. Estimates of the sensitivities of these missions to a SGWB, at the $\Omega_{\rm gw} \sim 10^{-3}$ -- $10^{-6}$ level, are given in Refs. \cite{gaia00,Makarov10,Mignard10}.

VLBI radio interferometry is another method that can be used to detect the astrometric effects of a SGWB on distant sources. This method detects the same pattern as that discussed in this paper for visible astrometry, and differs from astrometry using the GAIA satellite in its longer duration (tens of years versus a few years for GAIA), and in the smaller number of sources, on the order of hundreds, that have currently been measured using this method. In the radio, the planned Square Kilometer Array (SKA) is also expected to be able to localize sources to within $\sim 10 \mu\text{as}$ \cite{Fomalont04}. Jaffe has estimated that with $10^6$ QSO sources, the SKA could achieve a sensitivity of order $\Omega_{\rm gw} \sim 10^{-6}$ \cite{Jaffe04}.

The astrometric signals due to a SGWB expected for a single object are quite small, on the order of $0.1 \: \mu$as yr$^{-1}$, much smaller than the typical intrinsic proper motion of a star in our Galaxy. We therefore propose to use quasars as our sources, since their extragalactic distances cause their expected intrinsic proper motions to be smaller than those expected from a SGWB \cite{gaia00}. The construction of a nonrotating reference frame using quasars in astrometric studies will remove the $l=1$ dipole component of the measured quasar proper motions, but will leave intact the $l=2$ and higher multipoles which are expected to be excited by GWs.

Using the estimate $N \sim 10^6$ (GAIA), $\Delta \theta \sim 10 \: \mu\text{as}$, $T \sim 1$ yr gives from Eq. (\ref{eqn:omapprox}) the estimate

\be \Omega_{\rm gw} \alt 10^{-6} \nonumber\ee

\noindent at $f \alt 10^{-8}$ Hz for astrometry. This is an interesting sensitivity level, roughly comparable with that obtainable with pulsar timing \cite{Jenet06}.

Astrometry has already been applied to obtain upper limits on $\Omega_{\rm gw}$ using a number of different observations.  First, Gwinn et. al analyzed limits on quasar proper motions obtained from VLBI astrometry, and obtained the upper limit $\Omega_{\rm gw} \alt 10^{-1}$ for $10^{-17} \, {\rm Hz} \alt f \alt 10^{-9} \, {\rm Hz}$ \cite{Gwinn97}. This limit was recently updated by Titov, Lambert and Gontier \cite{Titov10}.  Finally, Linder analyzed observed galaxy correlation functions to obtain the limit $\Omega_{\rm gw} \alt 10^{-3}$ for $10^{-16} \, {\rm Hz} \alt f \alt 10^{-10} \, {\rm Hz}$ \cite{Linder88}. 

All of these analyses used a relatively simple model of the effect of gravitational waves on proper motions. In this paper we give a detailed computation of the spectrum of angular fluctuations produced by a stochastic background, including the relative strengths of E- and B-type multipoles for each order $l$.  In a subsequent paper we will follow up with a derivation of the optimal data analysis method and a computation of the $\Omega_{\rm gw}$ sensitivity level, to confirm the existing crude estimates of the sensitivity of future astrometric missions such as GAIA.

\subsection{Summary of results}

For a source in the direction ${\bf n}$, the effect of the GW background is to produce an apparent angular deflection $\delta {\bf n}({\bf n},t)$. We first find a general formula for the angular deflection of a photon, for an arbitrary GW signal $h_{ij}$, emitted by a source that can be at a cosmological distance. This deflection is derived in Secs. \ref{sec:mink} and \ref{sec:FRW} below, and is given by [cf. Eq. (\ref{eqn:deflFRW})]

\begin{align} \delta n^i =& \frac{1}{2} \Bigg\{ n^j h_{ij}(0) - n^i n^j n^k h_{jk}(0)- \frac{\omega_0}{\zeta_s}\left( \delta^{ik} - n^i n^k \right)n^j \cdot \nonumber\\&\cdot \left[ -2 \int_0^{\zeta_s} d\zeta' \int_0^{\zeta'} d\zeta'' h_{jk,0}(\zeta'') + n^l \int_0^{\zeta_s} d\zeta' \int_0^{\zeta'} d\zeta'' \left( h_{jk,l}(\zeta'') + h_{kl,j}(\zeta'') - h_{jl,k}(\zeta'') \right) \right] \Bigg\}.\nonumber\end{align}

\noindent Here, $\mathbf{n}$ is the direction to the source, $\omega_0$ is the emitted frequency of the photon, $\zeta$ parametrizes the path of the photon $\tau(\zeta) = \tau_0 + \omega_0 \zeta$, $ x^i(\zeta) = -\zeta \omega_0 n^i$, $h_{ij}(\tau,\mathbf{x})$ is treated as a function of $\zeta$ through this parametrization of the photon path,  $\zeta_s$ is the value of $\zeta$ at the emission event of the photon at the source, and the spacetime metric is

\be ds^2 = a(\tau)^2 \left\{ -d\tau^2 + \left[\delta_{ij} + h_{ij}(\tau,\mathbf{x})\right] dx^i dx^j \right\}. \nonumber \ee

We then specialize to the limit in which the sources are many gravitational wavelengths away and to plane waves propagating in the direction $\mathbf{p}$ to obtain a simple formula, which generalizes a previous result of Pyne et al \cite{Pyne96}. We find that the deflection, as a function of time $\tau$ and direction on the sky $\mathbf{n}$, is given by

\be \delta n^{\hat{i}}(\tau,{\bf n}) = \frac{n^i + p^i}{2 (1 + \mathbf{p}\cdot\mathbf{n})} h_{jk}(0) n_j n_k - \frac{1}{2} h_{ij}(0) n_j,\nonumber\ee

\noindent where $\mathbf{p}$ is the direction of propagation of the GW, and $h_{ij}(0)$ is the GW field evaluated at the observer, $h_{ij}(\tau,{\bf 0})$.

The main result of this paper is a computation of the statistical properties of the angular deflection resulting from a SGWB, which is carried out in Secs. \ref{sec:corr} and \ref{sec:spect}. The apparent angular deflection caused by such a GW background is a stationary, zero-mean, Gaussian random process.  We compute the fluctuations in $\delta {\bf n}$ by making two different approximations: (i) The GW modes which contribute to the deflection have wavelengths $\lambda$ which are short compared to the horizon size $c \, H_0^{-1}$ today. (ii) The mode wavelengths $\lambda$ are short compared to the distances to the sources; this same approximation is made in pulsar timing searches for GWs \cite{Detweiler79}. Since our calculations are only valid for GWs with wavelengths much smaller than the horizon, the contribution from waves with wavelengths comparable to the horizon scale will cause a small deviation from our results (on the order of a few percent for a white GW spectrum).

The total power in angular fluctuations is then

\be \left< \delta {\bf n}({\bf n},t)^2 \right> = \theta_{\rm rms}^2 = \frac{1}{4 \pi^2} \int d\ln f \left( \frac{H_0}{f} \right)^2 \Omega_{\rm gw}(f). \label{eqn:2ptnn} \ee

\noindent Taking a time derivative gives the spectrum of fluctuations of angular velocity or proper motion:

\be \left< \delta \dot{\bf n}({\bf n},t)^2 \right> = \int d\ln f H_0^2 \Omega_{\rm gw}(f), \nonumber \ee

\noindent which gives a rms angular velocity $\omega_{\rm rms}(f)$ of order 

\be \omega_{\rm rms}(f) \sim H_0 \sqrt{\Omega_{\rm gw}} \sim 10^{-2} \mu \text{as} \: \text{yr}^{-1} \left( \frac{\Omega_{\rm gw}}{10^{-6}} \right)^{1/2}. \nonumber\ee

\noindent This is the signal that we hope to detect.

\bigskip
We now discuss how the angular fluctuations are distributed on different angular scales, or equivalently how the power is distributed in the spherical harmonic index $l$. The total angular fluctuations can be written as

\be \left< \delta {\bf n}({\bf n},t)^2 \right> = \int d\ln f \sum_{l=2}^\infty \left[ \theta_{{\rm rms}, l}^{E}(f)^2 + \theta_{{\rm rms}, l}^{B}(f)^2 \right]. \label{eqn:thrms}\ee

\noindent Here $\theta^{E}_{{\rm rms},l}(f)^2$ is the total electric-type power in angular fluctuations per unit logarithmic frequency in multipole sector $l$, and $\theta^{B}_{{\rm rms},l}(f)^2$ is the corresponding magnetic-type power.  These quantities can be written as

\be \theta_{{\rm rms}, l}^{Q}(f)^2 = \theta_{\rm rms}^2 \,g_{Q} \,\sigma(f) \, \alpha^{QQ}_l, \ee

\noindent where $Q=E$ or $B$. The various factors in this formula are as follows.  The factors $g_{E}$ and $g_{B}$ are the fractions of the total power carried by E-modes and B-modes, respectively, and satisfy $g_{E} + g_{B}=1$.  Their values are $g_{E} = g_{B} = 1/2$, implying that electric- and magnetic-type fluctuations have equal power. The function $\sigma(f)$ describes how the power is distributed in frequency, and is the same for all multipoles, both electric and magnetic.  It is normalized so that $\int d\ln f \sigma(f) =1$, and is given explicitly by [cf.\ Eq.\ (\ref{eqn:2ptnn}) above]

\be \sigma(f) = \frac{ f^{-2} \: \Omega_{\rm gw}(f)}{\int d\ln f' f^{\prime    \, -2} \: \Omega_{\rm gw}(f')}. \label{eqn:sigma} \ee

\begin{figure}[t!]
\centering
\includegraphics[width=0.6\textwidth]{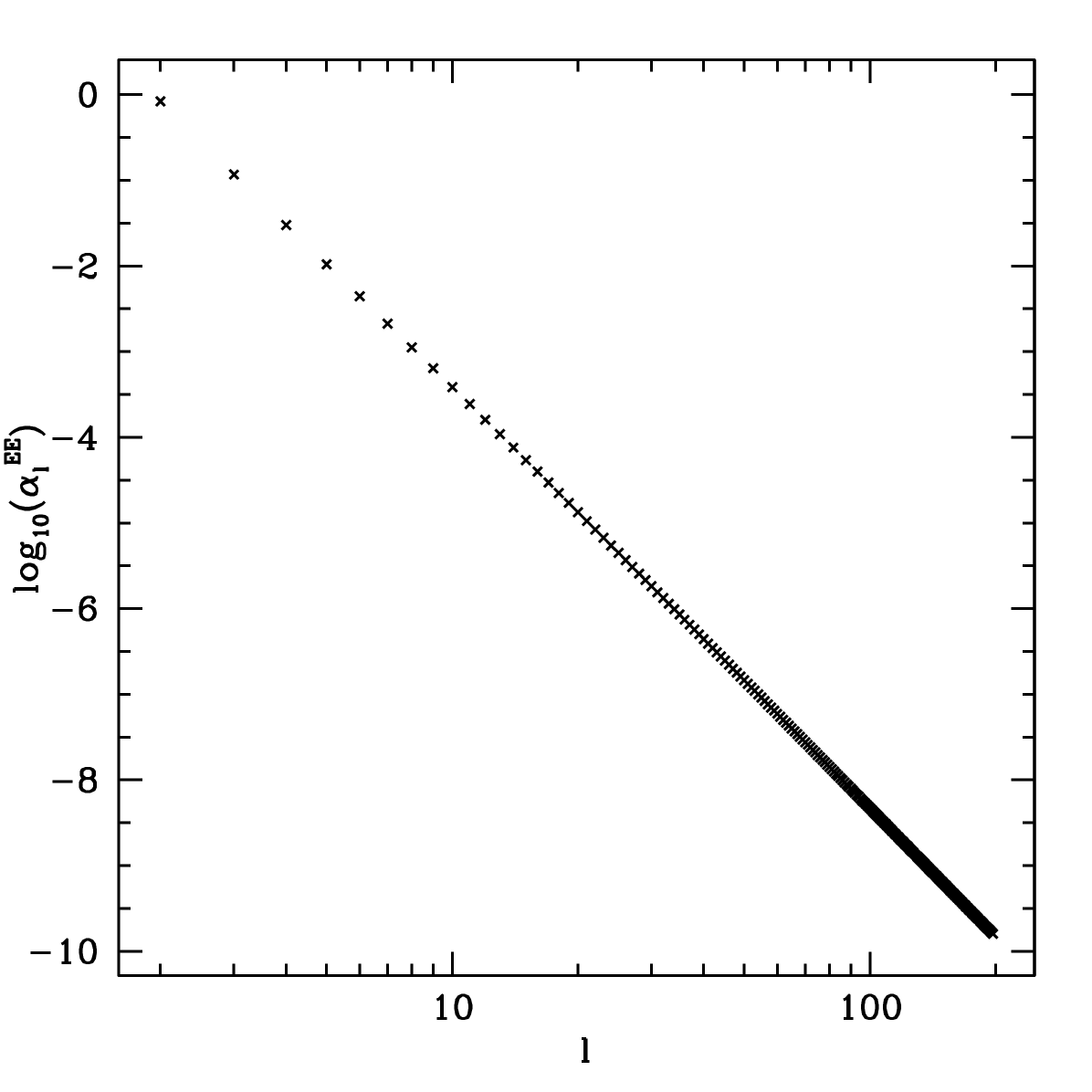}
\caption{Here we plot the coefficients $\alpha_l^{EE}$ as defined in Eq. (\ref{eqn:SQlf})  vs multipole $l$.}
\label{fig:alphaQ}
\end{figure}

\noindent Finally, the angular spectra $\alpha^{EE}_l$ and $\alpha^{BB}_l$ describe how the power is distributed in different multipoles, starting with the quadrupole at $l=2$, and are normalized so that

\be \sum_{l=2}^\infty \alpha^{QQ}_l =1 \ee

\noindent for $Q=E$ and $Q=B$.  We show that $\alpha^{EE}_l = \alpha^{BB}_l$, and this spectrum is plotted in Fig. \ref{fig:alphaQ} and tabulated in table \ref{tab:alphaQ}. These coefficients are well fit by the power law $\alpha^{EE}_l = 32.34 \: l^{-4.921}$. We note that the result for the quadrupole, $\alpha^{EE}_2 = 5/6$, has previously been derived using a different method in Ref. \cite{Pyne96}.

\begin{table*}
\centering
\caption{ First 10 nonzero multipole coefficients $\alpha_l^{EE}$ as defined in Eq. (\ref{eqn:SQlf}) and plotted in Fig. \ref{fig:alphaQ}.\label{tab:alphaQ} }
\begin{tabular}{ | l | r | }
\hline
\begin{centering} $l$ \end{centering}& \multicolumn{1}{c|}{$\alpha_l^{EE}$}\\
\hline
\hline
2 & 0.833 333 \\
3 & 0.116 667 \\
4 & 0.03 \\
5 & 0.010 476 2 \\
6 & 0.004 421 77 \\
7 & 0.002 125 85 \\
8 & 0.001 124 34 \\
9 & 0.000 639 731 \\
10 & 0.000 385 675 \\
11 & 0.000 243 696 \\
\hline
\end{tabular}
\end{table*}

\section{Calculation of Astrometric Deflection in a Minkowski Background Spacetime}  \label{sec:mink}

\subsection{ Setting the stage---Minkowski Calculation }

We will first calculate the angular deflection due to a small GW perturbation on a flat background metric, 

\be ds^2 \equiv g_{\mu \nu} dx^{\mu} dx^{\nu} = -dt^2 + (\delta_{ij} + h_{ij})dx^i dx^j.\ee

\noindent We are considering the effect of these GWs on a photon traveling from a source to an observer, with an unperturbed worldline $x^{\alpha}_0(\lambda) = \omega_0 ( \lambda, - \lambda {\bf n})+ (t_0,0,0,0),$ where $-{\bf n}$ is the direction of the photon's travel, $\omega_0$ is its unperturbed frequency, and the photon is observed at the origin at time $t_0$. The photon's unperturbed 4-momentum is given by $k^{\alpha}_0 = \omega_0 (1, -{\bf n}).$ 

To calculate the geodesics that the photon, source and observer follow, we need the connection coefficients in this metric. There are three nonzero connection coefficients,

\be \Gamma^{k}_{0 i} = \frac{1}{2} h_{k i, 0},\quad \Gamma^{0}_{i j} = \frac{1}{2} h_{i j , 0},\quad \Gamma^{k}_{i j} = \frac{1}{2} [ h_{k i, j} + h_{k j, i} - h_{i j, k} ]. \label{eqn:gamma}\ee

\noindent First, using the geodesic equation

\be \frac{d^2x^{\alpha}}{d\tau^2} = - \Gamma^{\alpha}_{\beta \gamma} u^{\beta} u^{\gamma}, \label{eqn:geodesic}\ee

\noindent it is straightforward to verify that the paths of stationary observers in these coordinates are geodesics. Therefore we can assume that both the source and observer are stationary in these coordinates, with

\begin{align} &x_{obs}^i(t) = 0 \nonumber \\ 
&x_{s}^i(t) = x_{s}^i = \text{constant}. \nonumber\end{align} 

\noindent The affine parameter of the source is therefore

\be \lambda_s = - \frac{\left|{\bf x}_s\right|}{\omega_0}. \nonumber\ee

\subsection{ Photon Geodesic}

Next, we solve the geodesic equation (\ref{eqn:geodesic}) for the path of a photon traveling from the source to the observer in the perturbed metric. We write this path as the sum of contributions of zeroth and first order in $h$,

\be x^{\alpha}(\lambda) = x^{\alpha}_0(\lambda) + x^{\alpha}_1(\lambda). \ee

\noindent Similarly, the photon 4-momentum is

\be k^{\alpha}(\lambda) = k^{\alpha}_0(\lambda) + k^{\alpha}_1(\lambda). \ee

\noindent We note that the connection coefficients are all first order in $h$, so keeping only first-order terms, we will use only the unperturbed photon 4-momentum in the geodesic equation, yielding

\be \frac{d^2x_1^0}{d\lambda^2} = -\frac{\omega_0^2}{2} n^i n^j h_{ij,0},\ee
\be  \frac{d^2x_1^k}{d\lambda^2} = - \frac{\omega_0^2}{2} [ -2 n^i h_{ki,0} + n^i n^j \left( h_{ki,j} + h_{kj,i} - h_{ij,k} \right) ]. \ee

We now integrate the geodesic equation to obtain the perturbed photon 4-momentum and trajectory. The right-hand sides are evaluated along the photon's unperturbed path from $\lambda=0$ at the present time back to $\lambda$, since they are already first order in $h$. We define

\begin{align} \mathcal{I}_{ij}(\lambda) &= \int_0^{\lambda} d\lambda' h_{ij,0}(\lambda'), &\mathcal{J}_{ijk}(\lambda) &= \int_0^{\lambda} d\lambda' h_{ij,k}(\lambda'),\nonumber\\
  \mathcal{K}_{ij}(\lambda) &= \int_0^{\lambda} d\lambda' \int_0^{\lambda'} d\lambda'' h_{ij,0}(\lambda''), &\mathcal{L}_{ijk}(\lambda) &= \int_0^{\lambda} d\lambda' \int_0^{\lambda'} d\lambda'' h_{ij,k}(\lambda''),\label{eqn:termdef}\end{align}
  
\noindent where $h_{ij}(\lambda)$ means $h_{ij}(t_0 + \omega_0 \lambda, -\omega_0 \lambda {\bf n})$. We find

\begin{align} k^{0}_1(\lambda) &= -\frac{\omega_0^2}{2} n^i n^j \mathcal{I}_{ij}(\lambda) + I_0, &k^{j}_1(\lambda) &= - \frac{\omega_0^2}{2} n^i R_{ij}+ J_0^j, \nonumber\\
x^{0}_1(\lambda) &= -\frac{\omega_0^2}{2} n^i n^j \mathcal{K}_{ij}(\lambda) + I_0 \lambda + K_0, &x^{j}_1(\lambda) &= - \frac{\omega_0^2}{2} n^i S_{ij} + J_0^j \lambda + L_0^j, \label{eqn:intkx}\end{align}

\noindent where $I_0$, $J_0^j$, $K_0$ and $L_0^j$ are constants of integration, and we have defined the quantities

\be R_{ij}(\lambda) \equiv \left[ -2 \mathcal{I}_{ij}(\lambda) + n^k \left( \mathcal{J}_{ijk}(\lambda) + \mathcal{J}_{jki}(\lambda) - \mathcal{J}_{ikj}(\lambda) \right) \right], \label{eqn:Rdef} \ee

\be S_{ij}(\lambda) \equiv \left[ -2 \mathcal{K}_{ij}(\lambda) + n^k \left( \mathcal{L}_{ijk}(\lambda) + \mathcal{L}_{jki}(\lambda) - \mathcal{L}_{ikj}(\lambda) \right) \right].  \ee

\subsection{ Boundary conditions }

We determine the eight integration constants $I_0$, $J_0^j$, $K_0$, and $L_0^j$ using the boundary conditions of the problem, namely, that the photon path passes through the detection event $x^{\mu}_{obs} = (t_0,0,0,0)$, that it is null, that the photon is emitted with the unperturbed frequency $\omega_0$, and that the photon path intersects the path of the source at some earlier time.

\begin{enumerate}

\item {\em Photon path must pass through detection event.}

First, the perturbed photon trajectory must pass through the detection event $t = t_0$, $x^i = 0$. Therefore,

\be x^{\mu}(0) = x^{\mu}_0(0) + x^{\mu}_1(0) = (t_0,0,0,0), \nonumber\ee

\noindent giving

\be K_0 = \frac{\omega_0^2}{2} n^i n^j \mathcal{K}_{ij}(0)=0, \quad L^j_0 = \frac{\omega_0^2}{2} n^i S_{ij}(0)=0, \ee

\noindent where we have used the fact that by definition $\mathcal{K}_{ij}(0)=S_{ij}(0)=0$.

\item {\em Photon geodesic is null.}

The geodesic of the photon must be null, which gives one more constraint, $g_{\mu \nu} k^{\mu} k^{\nu} = 0$. This is already true to zeroth order. To first order we get

\be 0 = h_{\mu \nu} k^{\mu}_0 k^{\nu}_0 + 2 \eta_{\mu \nu} k^{\mu}_1 k^{\nu}_0, \nonumber\ee 

\noindent where $k^{\alpha}_0 = \omega_0 (1, -{\bf n})$. Inserting the expression for the perturbed 4-momentum $k_1^{\alpha}$ given by Eqs. (\ref{eqn:termdef}) , (\ref{eqn:intkx}) and (\ref{eqn:Rdef}) , and simplifying using

\be \frac{\text{d}}{\text{d}\lambda} h_{ij} = \omega_0 h_{ij,0} - \omega_0 n_k h_{ij,k} \label{eqn:dh}\ee

\noindent shows that  all of the terms involving $\lambda$ cancel out, as they must, leaving the condition

\be I_0 + n_i J_0^i = \frac{1}{2}\omega_0 n^i n^j h_{ij}(0). \label{eqn:nullcnst}\ee

\item {\em Photon is emitted with frequency $\omega_0$.}

The photon is emitted at the source with the unperturbed frequency $\omega_0 = - g_{\mu \nu} k^{\mu} u_s^{\nu}$. The 4-velocity of the source is $u_s^{\mu} = (1, 0, 0, 0)$ as it has constant spatial coordinate position, so the constraint becomes $-g_{\mu 0} k^{\mu} = \omega_0$. The source emits the photon at $\lambda = \lambda_s$, so from Eq. (\ref{eqn:intkx}) this yields

\be I_0 = \frac{\omega_0^2}{2} n^i n^j \mathcal{I}_{ij}(\lambda_s) \label{eqn:w0cnst}\ee

\item {\em Perturbed photon path must hit source worldline somewhere.}

The constraint that the perturbed photon trajectory must hit the source worldline somewhere can be written as 

\be x^j(\tilde{\lambda_s}) = x_s^j = x^j_0(\tilde{\lambda_s}) + x^j_1(\tilde{\lambda_s})\ee

\noindent for some $\tilde{\lambda_s}$. To zeroth order we have $\tilde{\lambda_s} = \lambda_s$, but there will be a first-order correction. Inserting the expression (\ref{eqn:intkx}) for the perturbation of the geodesic gives

\be x_s^j = -\omega_0 \tilde{\lambda_s} n^j - \frac{\omega_0^2}{2} n^i S_{ij}(\tilde{\lambda_s}) + \tilde{\lambda_s} J_0^j. \label{eqn:source} \ee

Projecting this equation perpendicular to ${\bf n}$ gives a formula for the perpendicular component of $J_0^i$,

\be J^i_{0\;\perp} = \frac{\omega_0^2}{2 \lambda_s} \left( \delta^{ik} - n^i n^k \right) n^j S_{jk}(\lambda_s). \ee

\noindent Here on the right-hand side we have replaced $\tilde{\lambda_s}$ with $\lambda_s$, which is valid to linear order. Adding to this our earlier result for the component of $J_0^i$ parallel to ${\bf n}$ in Eqs. (\ref{eqn:nullcnst}) and (\ref{eqn:w0cnst}) gives

\be J_0^i = \frac{\omega_0^2}{2 \lambda_s} n^j S_{jk}(\lambda_s) \left( \delta^{ik} - n^i n^k \right) -  \frac{\omega_0^2}{2} n^i n^j n^k \mathcal{I}_{jk}(\lambda_s) + \frac{1}{2} \omega_0 n^i n^j n^k h_{jk}(0). \ee

\end{enumerate}

\subsection{ Perturbation to Observed Frequency }\label{sec:omegacomp}

We calculate the observed photon frequency $\omega_{obs} = -g_{\mu \nu} k^{\mu} u_{obs}^{\nu}$, where $u_{obs}^{\nu} = (1,0,0,0)$, and check our result against standard formulas for the frequency shift, used in pulsar timing searches for GWs \cite{Anholm09}. The observed frequency is, from Eqs. (\ref{eqn:intkx}) and (\ref{eqn:w0cnst}),

\be \omega_{obs} = k^0(0) = \omega_0 + I_0 = \omega_0 + \frac{\omega_0^2}{2} n^i n^j \mathcal{I}_{ij}(\lambda_s). \label{eqn:obsfreq}\ee

\noindent Using the definition (\ref{eqn:termdef}), the perturbed redshift is therefore

\be z\equiv \frac{\omega_0 - \omega_{obs}}{\omega_0} = -\frac{\omega_0}{2} n^i n^j \int_0^{\lambda_s} d\lambda' h_{ij,0}(\lambda'). \label{eqn:z}\ee

\noindent For a plane wave traveling in the direction of the unit vector ${\bf p}$, we have

\be h_{ij} = h_{ij}(t-\mathbf{p}\cdot\mathbf{x})=h_{ij}\left[\omega_0 \lambda (1+ \mathbf{p}\cdot\mathbf{n})\right], \nonumber\ee

\noindent giving

\be h_{ij,0}\equiv\frac{\partial}{\partial t} h_{ij}=\frac{1}{\omega_0 (1+\gamma)} \frac{\partial}{\partial \lambda} h_{ij}, \nonumber\ee

\noindent where $\gamma = \mathbf{p}\cdot\mathbf{n}$. This gives for the redshift

\be z = -\frac{1}{2 (1+\gamma)} n^i n^j \left[ h_{ij}(\lambda_s) - h_{ij}(0) \right], \ee

\noindent which agrees with \cite{Anholm09} up to a sign, which is an error in their calculation \cite{CPC}.

\subsection{ Local Proper Reference Frame of Observer }

We must also account for the changes induced in the basis vectors of the observer's local proper reference frame due to the presence of the GW. We introduce a set of orthonormal basis vectors $\vec{e}_{\hat{\alpha}}$ which are parallel transported along the observer's worldline, with $\vec{e}_{\hat{0}} = \vec{u}$. The parallel transport equation for the spatial vectors gives

\be u^{\alpha} e^{\beta}_{\hat{j} ; \alpha} = u^{\alpha} \left[\partial_{\alpha} e^{\beta}_{\hat{j}} + \Gamma^{\beta}_{\alpha \gamma} e^{\gamma}_{\hat{j}}\right] = 0. \label{eqn:plltrans}\ee

\noindent We separate the basis vectors into two pieces, $ e^i_{\hat{j}} = \delta^i_{\hat{j}} + \delta e^i_{\hat{j}}$, where we assume that the unperturbed basis vectors are aligned with the coordinate basis directions.

Using ${\vec u} = \partial_t$, and the connection coefficients (\ref{eqn:gamma}) of the metric , Eq. (\ref{eqn:plltrans}) gives us an expression for the perturbation to the basis tetrad,

\be \delta e^i_{\hat{j}}(t) = - \frac{1}{2} h_{i\hat{j}}(t,{\bf 0}) + \omega_{i \hat{j}}, \nonumber\ee

\noindent where $\omega^i_{\hat{j}}$ is a matrix of constants. Now, we observe that $e_{\hat{j}}$ is an orthonormal set of three-vectors, which gives us six constraints on the constants $\omega^i_{\hat{j}}$,

\be \left( \eta_{mn} + h_{mn} \right)\left( \delta^m_{\hat{j}} + \delta e^m_{\hat{j}} \right) \left( \delta^n_{\hat{k}} + \delta e^n_{\hat{k}} \right) = \delta_{\hat{j} \hat{k}}.\nonumber\ee

\noindent This is identically correct to zeroth order; to first order we get $ \delta e_{j \hat{k}} + \delta e_{k \hat{j}} + h_{jk} = 0$ or, inserting our equation for $\delta e$, and assuming that $h_{ij} = h_{ji}$, we find $\omega_{ij} = -\omega_{ji}$, i.e. that the constants $\omega_{ij}$ are antisymmetric in their indices. These constants parametrize an arbitrary infinitesimal time-independent rotation. Evaluating now at the detection event gives

\be \delta e^i_{\hat{j}} = - \frac{1}{2} h_{i\hat{j}}(0) + \omega_{i\hat{j}}. \ee

\noindent For the remainder of this paper we will set to zero the term $\omega_{i\hat{j}}$, since it corresponds to a time-independent, unobservable angular deflection. The deflections caused by GWs will be observable because of their time dependence.

\subsection{ Observed Angular Deflection }

We can express the 4-momentum of the incoming photon in the above reference frame as

\be k^{\alpha}(0) = \omega_{obs} u^{\alpha} - \omega_{obs} n^{\hat{j}} e^{\alpha}_{\hat{j}}, \label{eqn:obsk}\ee

\noindent where $\delta_{\hat{j} \hat{k}} n^{\hat{j}} n^{\hat{k}} = 1$, $u^{\alpha}$ is the observer's 4-velocity, and $\omega_{obs}$ is given by Eq. \ref{eqn:obsfreq}. Note that we evaluate all quantities at the detection event $t = t_0$, ${\bf x} = 0$. Plugging in our results for the perturbed 4-momentum and the observed frequency, we obtain an equation for the observed direction to the source $n^{\hat{j}}$

\begin{align} k^i(0) =&  - \omega_0 n^i + \frac{\omega_0^2}{2 \lambda_s} n^j S_{jk}(\lambda_s) \left( \delta^{ik} - n^i n^k \right) -  \frac{\omega_0^2}{2} n^i n^j n^k \mathcal{I}_{jk}(\lambda_s) + \frac{1}{2} \omega_0 n^i n^j n^k h_{jk}(0)\nonumber\\
 =& - \left(\omega_0 + \frac{\omega_0^2}{2} n^k n^l \mathcal{I}_{kl}(\lambda_s) \right) n^{\hat{j}} \left( \delta^i_j - \frac{1}{2} h^i_j(0) \right). \label{eqn:obsdir}\end{align}

We decompose the direction to the source into zeroth and first-order pieces as $n^{\hat{j}} = n^{\hat{j}}_0 + \delta n^{\hat{j}}$. The zeroth-order terms in Eq. (\ref{eqn:obsdir}) give us $n_0^{\hat{j}} = n^j$. Plugging this into the first order terms and simplifying, we find the perturbation to the source direction

\be \delta n^{\hat{i}} = \frac{1}{2} \left\{n^j h_{ij}(0) - \frac{\omega_0}{\lambda_s} n^j S_{jk}(\lambda_s) \left( \delta^{ik} - n^i n^k \right) - n^i n^j n^k h_{jk}(0) \right\}.\nonumber\ee

\noindent Inserting our definition of $S_{jk}$, we obtain the solution to the source direction perturbation in Minkowski space

\begin{align} \delta n^{\hat{i}} =& \frac{1}{2} \Bigg\{ n^j h_{ij}(0) - n^i n^j n^k h_{jk}(0) - \frac{\omega_0}{\lambda_s}\left( \delta^{ik} - n^i n^k \right)n^j \nonumber\\&\times \left[ -2 \int_0^{\lambda_s} d\lambda' \int_0^{\lambda'} d\lambda'' h_{jk,0}(\lambda'') + n^l \int_0^{\lambda_s} d\lambda' \int_0^{\lambda'} d\lambda'' \left( h_{jk,l}(\lambda'') + h_{kl,j}(\lambda'') - h_{jl,k}(\lambda'') \right) \right] \Bigg\}. \label{eqn:defl}\end{align}

\noindent As a check of the calculation, we see that $\delta n^{\hat{i}}$ is orthogonal to $n^i$, so that $n^i + \delta n^{\hat{i}}$ is a unit vector, as expected.

We now specialize to the case of a plane wave propagating in the direction of the unit vector ${\bf p}$,

\be h_{ij}(t,\mathbf{x}) = h_{ij}(t-\mathbf{p}\cdot\mathbf{x}). \nonumber\ee

\noindent Using the identity (\ref{eqn:dh}) we can reduce the double integrals in Eq. (\ref{eqn:defl}) to single integrals, obtaining

\be \delta n^{\hat{i}} = \left( \delta^{ik} - n^i n^k \right) n^j \Bigg\{ - \frac{1}{2} h_{jk}(0) + \frac{p_k n_l}{2 (1 + \mathbf{p}\cdot\mathbf{n})} h_{jl}(0) + \frac{1}{\lambda_s} \int_0^{\lambda_s} d\lambda \left[ h_{jk}(\lambda) - \frac{p_k n_l}{ 2 (1 + \mathbf{p}\cdot\mathbf{n}) } h_{jl}(\lambda) \right] \Bigg\}. \label{eqn:defl1}\ee

\noindent Evaluating this explicitly for the plane wave

\be h_{ij}(t,\mathbf{x}) = \text{Re}\left[ \mathcal{H}_{ij} e^{-i \Omega (t - \mathbf{p}\cdot\mathbf{x})} \right] \nonumber\ee

\noindent gives

\begin{align} \delta n^{\hat{i}} =& \text{Re}\Bigg[ \Bigg( \left\{ 1 + \frac{i (2 + \mathbf{p}\cdot\mathbf{n}) }{\omega_0 \lambda_s \Omega (1 + \mathbf{p}\cdot\mathbf{n})} \left[1 - e^{-i \Omega \omega_0 (1 + \mathbf{p}\cdot\mathbf{n}) \lambda_s }\right] \right\} n^i \nonumber\\
&+ \left\{ 1 + \frac{i}{\omega_0 \lambda_s \Omega (1 + \mathbf{p}\cdot\mathbf{n})} \left[1 - e^{-i \Omega \omega_0 (1 + \mathbf{p}\cdot\mathbf{n}) \lambda_s }\right] \right\} p^i \Bigg) \frac{n^j n^k \mathcal{H}_{j k} \text{e}^{-i \Omega t_0}}{2 (1 + \mathbf{p}\cdot\mathbf{n})}\nonumber\\
&- \left\{ \frac{1}{2} + \frac{i}{\omega_0 \lambda_s \Omega (1 + \mathbf{p}\cdot\mathbf{n})} \left[1 - e^{-i \Omega \omega_0 (1 + \mathbf{p}\cdot\mathbf{n}) \lambda_s }\right] \right\} n^j \mathcal{H}^i_j \text{e}^{-i \Omega t_0}\Bigg] .
\label{eqn:m1}\end{align}

If we define the observed angles $(\theta,\phi)$ by $n^{\hat{i}} = \left(\sin\theta \cos\phi, \sin\theta \sin\phi, \cos\theta\right)$, then the observed angular deflections are

\be \delta \theta = \text{e}^{\hat{i}}_{\hat{\theta}} \delta n^{\hat{i}}, \quad \delta \phi = \frac{\text{e}^{\hat{i}}_{\hat{\phi}} \delta n^{\hat{i}}}{\sin\theta}, \label{eqn:sphbasis}\ee

\noindent where $\text{e}^{\hat{i}}_{\hat{\theta}} = \left(\cos\theta \cos\phi, \cos\theta \sin\phi, -\sin\theta\right)$ and $\text{e}^{\hat{i}}_{\hat{\phi}} = \left(-\sin\phi,\cos\phi,0\right)$.

As another check of our calculation, we now compare our result with the coordinate (gauge-dependent) angular deflection computed by Yoo et al. \cite{Yoo09}. Starting from our Eq. (\ref{eqn:defl}), we disregard the first term, which arises from the change from the coordinate basis to the parallel transported orthonormal basis. The remaining terms in Eq. (\ref{eqn:defl}) give the coordinate angular deflection $\delta n^i$. Simplifying using the identity (\ref{eqn:dh}) and the identity $\int_0^{x} dx' \int_0^{x'}dx'' f(x'') = \int_0^{x} dx' (x-x') f(x')$ gives

\be \delta n^i = - \frac{1}{2} n^i n^j n^k h_{jk}(0) + \left( \delta^{ij} - n^i n^j \right) \int^{\lambda_s}_0 d\lambda \left\{\frac{h^{jk}(\lambda) - h^{jk}(0)}{\lambda_s} n_k + \frac{\omega_0}{2} \left( \frac{\lambda_s - \lambda}{\lambda_s} \right) \partial_j \left( n^k n^l h_{kl} \right)\right\}. \label{eqn:Yoo}\ee

\noindent When combined with Eqs. (\ref{eqn:sphbasis}), this agrees with Eqs. (13) and (14) of \cite{Yoo09}, specialized to only tensor perturbations, up to an overall sign. The sign flip is due to the fact that Ref. \cite{Yoo09} uses a convention for the sign of angular deflection, explained after their Eq. (16), which is opposite to ours.

\subsection{ The Distant Source Limit }

We now specialize to the limit where the distance $\omega_0 \left|\lambda_s\right|$ to the source is large compared to the wavelength $\sim c \: \Omega^{-1}$ of the GWs. As discussed in the Introduction, astrometry is potentially sensitive to waves with a broad range of frequencies, extending from the inverse of the observation time (a few years) down to the Hubble frequency. Therefore this assumption is a nontrivial limitation on the domain of validity of our analysis. However, for sources at cosmological distances (the most interesting case), this assumption is not a significant limitation. 

In this limit, we can neglect the second term in each of the three small square brackets in Eq. (\ref{eqn:m1}), giving

\be \delta n^{\hat{i}}(t,\mathbf{n}) = \text{Re}\left[ \left( n^i + p^i \right) \frac{\mathcal{H}_{jk} n_j n_k \text{e}^{-i\Omega t}}{2 (1 + \mathbf{p}\cdot\mathbf{n})} - \frac{1}{2} \mathcal{H}_{ij} n_j \text{e}^{-i\Omega t} \right], \label{eqn:dsl3}\ee

\noindent where we have written $t$ for $t_0$. This result agrees with and generalizes a calculation of Pyne et al. \cite{Pyne96}. We note that this same approximation is used in pulsar timing searches for GWs \cite{Anholm09}. In that context the approximation is essentially always valid, since pulsar distances are large compared to a few light years, and the properties of pulsar frequency noise imply that that pulsar timing is only sensitive to GWs with periods of order the observation time, and not much lower frequencies, unlike the case for astrometry.

\section{Generalization to Cosmological Spacetimes}  \label{sec:FRW}

Of course, we do not live in Minkowski space. The apparent homogeneity and isotropy of the Universe imply that our Universe has a Friedmann-Robertson-Walker (FRW) geometry, with line element

\be ds^2 = g_{\alpha \beta} dx^{\alpha} dx^{\beta} = a(\tau)^2 \left\{ -d\tau^2 + \left[\delta_{ij} + h_{ij}(\tau,\mathbf{x})\right] dx^i dx^j \right\}, \label{eqn:FRW}\ee

\noindent where $\tau$ is conformal time, and we specialize to the transverse traceless gauge in which $\delta^{ij} h_{ij} = \delta^{ij} \partial_i h_{jk} =0$. To translate our calculation in Minkowski spacetime to this new metric, we define an unphysical, conformally related metric ${\bar g}_{\alpha\beta} = a(\tau)^{-2}
g_{\alpha\beta}$ given by

\be {\bar g}_{\alpha\beta} dx^\alpha dx^\beta = -d\tau^2 + \left[\delta_{ij} + h_{ij}(\tau,{\bf x}) \right] dx^i dx^j, \label{eqn:gbar}\ee

\noindent which has an associated unphysical derivative operator ${\bar \nabla}_\alpha$.  

\subsection{ Stationary Observers are Freely Falling }

As before, it is straightforward to check that observers who are stationary in the coordinates (\ref{eqn:FRW}) are freely falling. Therefore we assume as before that the observer and source are stationary:

\be x_{obs}^i(t) = 0, \quad x_{s}^i(t) = x_{s}^i. \nonumber\ee

\subsection{ Null Geodesic in the Conformal Metric }

Let us consider a photon traveling from a distant source to us, which follows a null geodesic in the physical metric $g_{\alpha \beta}$. Its path is also a null geodesic of the conformally related metric ${\bar g}_{\alpha\beta}$, though it is not affinely parametrized in this metric \cite{Wald84}. Specifically, the physical 4-momentum of the photon $k^{\mu}$ must satisfy the geodesic equation $k^{\mu} \nabla_{\mu} k_{\nu} = 0$. If we define a conformally related, unphysical 4-momentum ${\bar k}_{\mu} = k_{\mu}$, whose contravariant components are then related to those of the physical 4-momentum by

\be {\bar k}^{\mu} = {\bar g}^{\mu\nu} {\bar k}_{\nu} = a(\tau)^2 g^{\mu\nu} {\bar k}_{\nu} = a(\tau)^2 g^{\mu\nu} k_{\nu} = a(\tau)^2 k^{\mu}, \ee

\noindent then we find that

\be  {\bar k}^{\mu} {\bar \nabla}_{\mu} {\bar k}_{\nu}  = a(\tau)^2 k^{\mu} {\bar \nabla}_{\mu} k_{\nu}. \ee

\noindent From \cite{Wald84} we know that for any vector $v^{\alpha}$, and conformally related derivatives $\nabla_{\alpha}$ and ${\bar \nabla}_{\alpha}$, we have $\nabla_{\alpha} v_{\beta} = {\bar \nabla}_{\alpha} v_{\beta} - C^{\gamma}_{\alpha \gamma} v_{\gamma}$, where $C^{\gamma}_{\alpha \gamma} = 2 \delta^{\gamma}_{ (\alpha}\nabla_{\beta)} \ln a - g_{\alpha \beta} g^{\gamma \delta} \nabla_{\delta} \ln a$. Thus, we find 

\begin{align} {\bar k}^{\mu} {\bar \nabla}_{\mu} {\bar k}_{\nu} &= a(\tau)^2 k^{\mu} \nabla_{\mu} k_{\nu} + a(\tau)^2 k^{\mu} k_{\rho} \left( 2 \delta^{\rho}_{ (\mu}\nabla_{\nu)} \ln a - g_{\mu \nu} g^{\rho \sigma} \nabla_{\sigma} \ln a \right) \nonumber\\ &= a(\tau)^2 k^{\mu} \nabla_{\mu} k_{\nu} + a(\tau)^2 \left( k^{\rho} k_{\rho} \nabla_{\nu} \ln a + k^{\mu} k_{\nu} \nabla_{\mu} \ln a - k_{\nu} k^{\sigma} \nabla_{\sigma} \ln a  \right) \nonumber\\&= a(\tau)^2 k^{\mu} \nabla_{\mu} k_{\nu}, \end{align}

\noindent where to get the last line we have used that the geodesic is null. Therefore, if $k^{\mu}$ is a null geodesic of the physical metric $g_{\mu\nu}$, then ${\bar k}^{\mu}$ is a null geodesic of the conformally related metric ${\bar g}_{\mu\nu}$. If $\lambda$ is an affine parameter of the geodesic, it will not be an affine parameter of the geodesic in the unphysical metric. The affine parameter ${\bar \lambda}$ in the unphysical metric is related to $\lambda$ by

\be \frac{d{\bar \lambda}}{d\lambda} = \frac{1}{a(\tau(\lambda))^2}. \ee

\subsection{ Parallel Transport of Basis Vectors in FRW Background Spacetime }

We next investigate the parallel transport of the observer's basis tetrad in a FRW background spacetime. From the form (\ref{eqn:FRW}) of the metric, we anticipate that the basis vectors must scale as $a^{-1}$ to remain normalized. Thus, we will define the basis vectors and their perturbations as 

\be e^i_{\hat{j}} = \frac{1}{a}\left(\delta^i_{\hat{j}} + \delta e^i_{\hat{j}}\right). \label{eqn:plfrw}\ee

\noindent The relevant connection coefficients are

\be \Gamma^i_{0k} = \frac{\dot{a}}{a} \delta^i_k + \frac{1}{2}\delta^{im} h_{mk,0}. \label{eqn:cc}\ee

\noindent The parallel transport equation (\ref{eqn:plltrans}) for the spatial basis vectors gives us

\be \partial_0 e^i_{\hat{j}} + \Gamma^i_{0 k} e^k_{\hat{j}} = 0. \ee

\noindent Plugging in the connection coefficients (\ref{eqn:cc}) and the basis vector expansion (\ref{eqn:plfrw}), we get

\be \partial_0 \delta e^i_{\hat{j}} + \frac{1}{2}\delta^{im} h_{m\hat{j}}=0, \ee

\noindent the same equation as before. The solution, as before, will be

\be \delta e^i_{\hat{j}}(t) = - \frac{1}{2} h^i_j(t). \ee

\subsection{ Generalization of Angular Deflection Computation }

We parametrize the photon path in the background spacetime by 

\be \tau(\zeta) = \tau_0 + \omega_0 \zeta, \quad x^i(\zeta) = -\zeta \omega_0 n^i, \ee

\noindent where $\zeta$ is an affine parameter of the unphysical metric (\ref{eqn:gbar}) (denoted ${\bar \lambda}$ above). From the decomposition (\ref{eqn:obsk}), the observed source direction is 

\be n^{\hat j} = \frac{g_{\alpha \beta} k^{\alpha} e^{\beta}_{\hat{j}}}{g_{\alpha \beta} k^{\alpha} u^{\beta}}. \ee

\noindent We rewrite all the quantities in this expression in terms of their conformally transformed versions

\be {\bar g}_{\alpha \beta} = a^{-2} g_{\alpha \beta}, \quad {\bar k}^{\alpha} = a^2 k^{\alpha}, \quad {\bar u}^{\alpha} = a u^{\alpha}, \quad  {\bar e}^{\alpha}_{\hat{j}} = a e^{\alpha}_{\hat{j}}, \ee

\noindent which are the quantities that are used in the Minkowski spacetime calculation of Sec. \ref{sec:mink}. This gives

\be n^{\hat j} = \frac{{\bar g}_{\alpha \beta} {\bar k}^{\alpha} {\bar e}^{\beta}_{\hat{j}}}{{\bar g}_{\alpha \beta} {\bar k}^{\alpha} {\bar u}^{\beta}}, \ee

\noindent the same expression as in Minkowski spacetime. Therefore, the final result is the same expression (\ref{eqn:defl}) as before, except that it is written in terms of the nonaffine parameter $\zeta$,

\begin{align} \delta n^{\hat i} =& \frac{1}{2} \Bigg\{ n^j h_{ij}(0) - n^i n^j n^k h_{jk}(0)- \frac{\omega_0}{\zeta_s}\left( \delta^{ik} - n^i n^k \right)n^j \nonumber\\
&\times \left[ -2 \int_0^{\zeta_s} d\zeta' \int_0^{\zeta'} d\zeta'' h_{jk,0}(\zeta'') + n^l \int_0^{\zeta_s} d\zeta' \int_0^{\zeta'} d\zeta'' \left( h_{jk,l}(\zeta'') + h_{kl,j}(\zeta'') - h_{jl,k}(\zeta'') \right) \right] \Bigg\}. \label{eqn:deflFRW}\end{align}

\subsection{ The Distant Source Limit }

We now specialize again to the limit where the distance to the source is large compared to the wavelength $\sim c \: \Omega^{-1}$ of the GWs. We also assume that the wavelength $c \: \Omega^{-1}$ is small compared to the horizon scale, but we allow the sources to be at cosmological distances.

Starting from Eq. (\ref{eqn:deflFRW}) and paralleling the derivation of Eq. (\ref{eqn:Yoo}) we obtain

\be \delta n^{\hat i}(\tau_0,{\bf n}) = \frac{1}{2} s_{ik} n_j h_{jk}(0) + \frac{s_{ik} n_j}{\zeta_s} \int_0^{\zeta_s} d\zeta \left[ h_{jk}(\zeta) - h_{jk}(0) \right] + \frac{\omega_0 s_{ik}}{2} \int_0^{\zeta_s} d\zeta \left( \frac{\zeta_s - \zeta}{\zeta_s} n_j n_l h_{jl,k}(\zeta) \right), \label{eqn:dsl1}\ee

\noindent where $s_{ik} = \delta_{ik} - n_i n_k$. Now the wave equation satisfied by the metric perturbation is 

\be \left[ \partial^2_{\tau} + 2 \frac{a_{,\tau}}{a} \partial_{\tau} - {\bf \nabla}^2 \right] h_{ij}(\tau,{\bf x}) = 0, \nonumber\ee

\noindent and plane wave solutions are of the form

\be h_{ij}(\tau,{\bf x}) = \text{Re} \left\{ \mathcal{H}_{ij} e^{i\Omega \mathbf{p}\cdot\mathbf{x}} q_{\Omega}(\tau) \right\}, \nonumber\ee

\noindent where the mode function $q_{\Omega}$ satisfies

\be q_{\Omega}'' + 2 \frac{a'}{a} q_{\Omega}' + \Omega^2 q_{\Omega} = 0. \label{eqn:dsl2}\ee

We now evaluate the angular deflection (\ref{eqn:dsl1}) for such a plane wave, in the limit where $\varepsilon \equiv a'/(\Omega a) \ll 1$, i.e. the limit where the wavelength $\sim a/\Omega$ of the GW is much smaller than the horizon scale $\sim a^2/a'$. In the second term in (\ref{eqn:dsl1}), the term $h_{jk}(\zeta)$ is rapidly oscillating, and so its integral can be neglected compared to the integral of $h_{jk}(0)$; corrections will be suppressed by powers of $\varepsilon$. In the third term in (\ref{eqn:dsl1}), the integrand is rapidly oscillating, and so the integral will be dominated by contributions near the endpoints, up to $\mathcal{O}(\varepsilon)$ corrections. However the integrand vanishes at $\zeta=\zeta_s$, and thus the integral is dominated by the region near $\zeta=0$. In that region we can use the leading order Wentzel-Kramers-Brillouin (WKB) approximation to the mode function solution of (\ref{eqn:dsl2}),

\be q_{\Omega}(\tau) = \frac{1}{a(\tau)} e^{-i \Omega \tau}, \nonumber\ee

\noindent and to a good approximation we can replace $a(\tau)$ by $a(\tau_0)$. Thus we see that the same answer is obtained for distant sources as in our Minkowski spacetime calculation, even for sources at cosmological distances. From Eq. (\ref{eqn:dsl3}) we obtain

\be \delta n^{\hat{i}}(\tau_0,{\bf n}) = \frac{n^i + p^i}{2 (1 + \mathbf{p}\cdot\mathbf{n})} h_{jk}(0) n_j n_k - \frac{1}{2} h_{ij}(0) n_j\label{eqn:dsl4}\ee

\noindent for plane waves in the direction ${\bf p}$.

\section{Calculation of Angular Deflection Correlation Function}  \label{sec:corr}

Now that we have calculated the deflection of the observed direction to a distant source due to an arbitrary metric perturbation $h_{ij}$, we would like to determine the properties of the deflection produced by a SGWB, such as that produced by inflation. 

\subsection{ Description of SGWB as a Random Process }

In the distant source limit, the angular deflection (\ref{eqn:dsl4}) depends only on the GW field $h_{ij}$ evaluated at the location of the observer for each direction of propagation $\mathbf{p}$. Moreover, we have restricted attention to modes with wavelengths short compared to the Hubble time. Therefore, it is sufficient to use a flat spacetime mode expansion to describe the stochastic background. This expansion is (see, e.g. Ref. \cite{Flanagan93})

\be h_{ij}(\mathbf{x}, t) = \sum_{A=+,\times} \int_0^{\infty} df \int d^2\Omega_{\mathbf{p}} \: h_{A \mathbf{p}}(f) \: e^{2 \pi i f (\mathbf{p}\cdot\mathbf{x} - t)} \: e^{A,\mathbf{p}}_{ij} + c.c., \label{eqn:hSGWB}\ee

\noindent where $f$ and $\mathbf{p}$ are the frequency and direction of propagation of individual GW modes, $h_{A\mathbf{p}}$ are the stochastic amplitudes of modes with polarization $A$ and direction $\mathbf{p}$, and the polarization tensors $e^{A,\mathbf{p}}_{ij}$ are normalized such that $e^{A,\mathbf{p}}_{ij} e^{B,\mathbf{p}*}_{ij} = 2 \delta^{AB}$. 

We will assume that $h_{ij}(\mathbf{x}, t)$ is a Gaussian random process, as it is likely to be the sum of a large number of random processes. We also assume that it is zero-mean and stationary. It follows that the mode amplitudes $h_{A \mathbf{p}}(f)$ satisfy

\begin{align} \langle h_{A \mathbf{p}}(f) \: h_{B \mathbf{p'}}(f') \rangle\;\: =&\: 0,\nonumber\\ \langle h_{A \mathbf{p}}(f) \: h_{B \mathbf{p'}}(f')^* \rangle =&\: \frac{3 H_0^2\Omega_{\rm gw}(f)}{32 \pi^3 f^3} \: \delta(f-f') \: \delta_{AB} \: \delta^2(\mathbf{p},\mathbf{p}') \label{eqn:h2pt}\end{align}

\noindent for $f,f'\geq 0$, where $H_0$ is the Hubble parameter and $\delta^2(\mathbf{p},\mathbf{p}')$ is the delta function on the unit sphere (see, e.g., \cite{Flanagan93}).

Since the angular deflection $\delta\mathbf{n}(\mathbf{n},t)$ depends linearly on the metric perturbation, it will also be a stationary, zero-mean, Gaussian random process, whose statistical properties are determined by its two-point correlation function $\langle \delta n^i \delta n^j \rangle$. Specializing our expression (\ref{eqn:dsl4}) for the angular deflection to the form (\ref{eqn:hSGWB}) of the metric perturbation, we find

\be \delta n^i(\mathbf{n},t) = \sum_{A=+,\times} \int_0^{\infty} df \int d^2\Omega_{\mathbf{p}} \: h_{A \mathbf{p}}(f) \: e^{-2\pi i f t} \: \mathcal{R}_{ikl}(\mathbf{n},\mathbf{p}) \: e^{A,\mathbf{p}}_{kl} + c.c. \label{eqn:dn}, \ee

\noindent where

\be \mathcal{R}_{ikl}(\mathbf{n},\mathbf{p}) = \frac{1}{2} \left[ \frac{\left(n_i + p_i\right) n_k n_l}{1 + \mathbf{p}\cdot\mathbf{n}} - n_k \delta_{il} \right]. \label{eqn:R}\ee

\subsection{ Power Spectrum of the Astrometric Deflections of the SGWB }

So, we need only evaluate the two-point correlation function to gain full knowledge of the statistical properties of the angular deflection due to the SGWB. Writing out this quantity explicitly using Eq. (\ref{eqn:dn}),

\begin{align} \langle \delta n^i(\mathbf{n},t) \: \delta n^{j}(\mathbf{n}',t') \rangle &= \sum_{A,B=+,\times} \int_0^{\infty} df df' \int d^2\Omega_{\mathbf{p}} d^2\Omega_{\mathbf{p}'} \bigg\langle \left[h_{A \mathbf{p}}(f) \; e^{-2\pi i f t} \; \mathcal{R}_{ikl}(\mathbf{n},\mathbf{p}) \; e^{A,\mathbf{p}}_{kl} + c.c.\right] \nonumber \\
&\times \left[h_{B \mathbf{p}'}(f')^*\; e^{2\pi i f' t'} \; \mathcal{R}_{jrs}(\mathbf{n}',\mathbf{p}') \; \left(e^{B,\mathbf{p}'}_{rs}\right)^* + c.c.\right] \bigg\rangle.\label{eqn:firstcor}\end{align}

\noindent The average, which is an average over ensembles, acts only on the stochastic amplitudes $h_{A \mathbf{p}}$. Using the mode two-point function (\ref{eqn:h2pt}) in Eq. (\ref{eqn:firstcor}), we get the simplified result

\be  \langle \delta n^i(\mathbf{n},t) \: \delta n^j(\mathbf{n}',t') \rangle = \int_0^{\infty} df \frac{3 H_0^2}{32 \pi^3} f^{-3} \Omega_{\rm gw}(f) e^{-2\pi i f (t-t')} H_{ij}(\mathbf{n},\mathbf{n}') + c. c., \label{eqn:2ptH}\ee

\noindent where we have defined

\be H_{ij}(\mathbf{n},\mathbf{n}') = \sum_{A=+,\times} \int d^2\Omega_{\mathbf{p}} \mathcal{R}_{ikl}(\mathbf{n},\mathbf{p}) \; e^{A,\mathbf{p}}_{kl} \; \mathcal{R}_{jrs}(\mathbf{n}',\mathbf{p}) \; \left(e^{A,\mathbf{p}}_{rs}\right)^*. \label{eqn:H}\ee

\subsection{ Basis Tensors and their symmetries \label{sec:basis}}

We simplify the expression (\ref{eqn:H}) for $H_{ij}$ further using the identity

\be \sum_{A=+,\times} e^{A,\mathbf{p}}_{ij} \left( e^{A,\mathbf{p}}_{kl} \right)^* = 2 P_{ijkl}, \label{eqn:ident}\ee

\noindent where $P_{ijkl}$ is the projection tensor onto the space of traceless symmetric tensors orthogonal to $\mathbf{p}$, given by

\be 2 P_{ijkl} = \delta_{ik} \delta_{jl} + \delta_{il} \delta_{jk} - \delta_{ij} \delta_{kl} + p_i p_j p_k p_l - \delta_{ik} p_j p_l - \delta_{jl} p_i p_k - \delta_{il} p_j p_k - \delta_{jk} p_i p_l + \delta_{ij} p_k p_l + \delta_{kl} p_i p_j. \label{eqn:P}\ee

\noindent This gives 

\be H_{ij}(\mathbf{n},\mathbf{n}') = 2 \int d^2\Omega_{\mathbf{p}} \mathcal{R}_{ikl}(\mathbf{n},\mathbf{p}) P_{klrs} \mathcal{R}_{jrs}(\mathbf{n}',\mathbf{p}).\label{eqn:HR}\ee

Noting that the correlation function (\ref{eqn:2ptH}) is perpendicular to $\mathbf{n}$ on its first index and $\mathbf{n}'$ on its second, we can decompose it onto a basis of tensors with this property,

\be H_{ij}(\mathbf{n},\mathbf{n}') = \alpha(\mathbf{n},\mathbf{n}') A_i A_j + \beta(\mathbf{n},\mathbf{n}') A_i C_j + \gamma(\mathbf{n},\mathbf{n}') B_i A_j + \sigma(\mathbf{n},\mathbf{n}') B_i C_j, \label{eqn:decomp}\ee

\noindent for some scalar functions $\alpha$, $\beta$, $\gamma$ and $\sigma$. Here we have defined 

\be \mathbf{A} = \mathbf{n} \times \mathbf{n}', \quad \mathbf{B} = \mathbf{n} \times \mathbf{A}, \quad \mathbf{C} = -\mathbf{n}' \times \mathbf{A}. \label{eqn:ABC}\ee 

\noindent We can deduce from Eq. (\ref{eqn:HR}) that $H_{ij}(\mathbf{n},\mathbf{n}')^* = H_{ji}(\mathbf{n}',\mathbf{n})$. Noting that $A_i(\mathbf{n}',\mathbf{n}) = -A_i(\mathbf{n},\mathbf{n}')$, and $B_i(\mathbf{n}',\mathbf{n}) = -C_i(\mathbf{n},\mathbf{n}')$, this symmetry applied to the expansion (\ref{eqn:decomp}) gives

\be \alpha(\mathbf{n},\mathbf{n}')^* = \alpha(\mathbf{n}', \mathbf{n}), \quad \sigma(\mathbf{n},\mathbf{n}')^* = \sigma(\mathbf{n}',\mathbf{n}), \quad \beta(\mathbf{n},\mathbf{n}')^* = \gamma(\mathbf{n}',\mathbf{n}). \nonumber\ee

\noindent We see from Eq. (\ref{eqn:H}) that $H_{ij}$ transforms as  tensor under rotations. This implies that the functions $\alpha$, $\beta$, $\gamma$ and $\sigma$ must be invariant under rotations, and can only depend on the angle $\Theta$ between $\mathbf{n}$ and $\mathbf{n}'$. Thus, $ \alpha(\mathbf{n},\mathbf{n}') = \alpha(\mathbf{n}',\mathbf{n})=\alpha(\Theta)$ and so forth, so $\alpha$ and $\sigma$ must be real.

Next, we note that the expression (\ref{eqn:HR}) for $H_{ij}(\mathbf{n},\mathbf{n}')$ is invariant under the parity transformation $\mathbf{n}\rightarrow-\mathbf{n}$ and $\mathbf{n}'\rightarrow-\mathbf{n}'$. Looking then at the basis tensors, we see that $\mathbf{A}$ is invariant under this transformation, while $\mathbf{B}$ and $\mathbf{C}$ change sign. Thus, in order to insure that $H_{ij}$ is invariant, it can only have terms multiplying $A_i A_j$ and $B_i C_j$, so $\beta(\Theta)=0=\gamma(\Theta)$. 

Having taken the symmetries of the problem into consideration, we have found $H_{ij}$ to be of the form

\be H_{ij}(\mathbf{n},\mathbf{n}') = \alpha(\Theta) A_i A_j + \sigma(\Theta) B_i C_j. \label{eqn:decomp2}\ee

\subsection{ Solving the General Integral }

We can evaluate the coefficients in the expansion (\ref{eqn:decomp2}) of $H_{ij}$ by contracting it with the basis tensors,

\be A^i A^j H_{ij} = \sin^4(\Theta) \alpha(\Theta), \quad B^i C^j H_{ij} = \sin^4(\Theta) \sigma(\Theta). \nonumber\ee

Rewriting these using Eq. (\ref{eqn:HR}), we find 

\be \alpha(\Theta) = \frac{2}{\sin^4(\Theta)} \int d^2\Omega_{\mathbf{p}} A^i \mathcal{R}_{ikl}(\mathbf{n},\mathbf{p}) P_{klrs} A^j \mathcal{R}_{jrs}(\mathbf{n}',\mathbf{p})^*, \ee

\be \sigma(\Theta) = \frac{2}{\sin^4(\Theta)} \int d^2\Omega_{\mathbf{p}} B^i \mathcal{R}_{ikl}(\mathbf{n},\mathbf{p}) P_{klrs} C^j \mathcal{R}_{jrs}(\mathbf{n}',\mathbf{p})^*. \ee

To simplify the calculation, we define the quantities $\kappa = \mathbf{n}\cdot\mathbf{p}$, $\kappa' = \mathbf{n}'\cdot\mathbf{p}$, $\lambda = \mathbf{n}\cdot\mathbf{n}'$, $\mu = \mathbf{A}\cdot\mathbf{p}$, which satisfy $\mu^2 + \lambda^2 + \kappa^2 + \kappa'^2 = 1 + 2 \lambda \kappa \kappa'$. Using these definitions and the definition (\ref{eqn:R}) of $\mathcal{R}_{ikl}$, we can write

\begin{align} A^i \mathcal{R}_{ikl}(\mathbf{n},\mathbf{p}) &= \frac{1}{2} n_k \left( \frac{\mu n_l}{1 + \kappa} -A_l \right),  &A^j \mathcal{R}_{jrs}(\mathbf{n}',\mathbf{p}) &= \frac{1}{2} n'_r \left( \frac{\mu n'_s}{1 + \kappa'} -A_s \right), \nonumber\\
 B^i \mathcal{R}_{ikl}(\mathbf{n},\mathbf{p}) &=  \frac{1}{2} n_k \left( -\frac{\kappa' + \lambda}{1 + \kappa} n_l + n'_l \right),  &C^j \mathcal{R}_{jrs}(\mathbf{n}',\mathbf{p}) &=  \frac{1}{2} n'_r \left( -\frac{\kappa + \lambda}{1 + \kappa'} n'_s + n_s \right). \nonumber\end{align}

\noindent We can then rewrite our expressions for $\alpha$ and $\sigma$,

\be \alpha(\Theta) = \frac{1}{4 \sin^4(\Theta)} \int d^2\Omega_{\mathbf{p}} 2P_{klrs} n_k \left( \frac{\mu n_l}{1 + \kappa} -A_l \right) n'_r \left( \frac{\mu n'_s}{1 + \kappa'} -A_s \right), \label{eqn:aP}\ee

\be \sigma(\Theta) = \frac{1}{4 \sin^4(\Theta)} \int d^2\Omega_{\mathbf{p}} 2P_{klrs} n_k \left( -\frac{\kappa' + \lambda}{1 + \kappa} n_l + n'_l \right) n'_r \left( -\frac{\kappa + \lambda}{1 + \kappa'} n'_s + n_s \right) .\label{eqn:sP}\ee

\noindent Let us define two new variables $\nu^2 = (1-\kappa^2)$, $\nu'^2 = (1-\kappa'^2)$. Applying the definition (\ref{eqn:P}) of the projection tensor $P_{klrs}$, we can calculate the necessary contractions of $P_{klrs}$ for $\alpha$,

\begin{align} 2 P_{klrs} n_k A_l n'_r A_s &=  \left( \lambda - \kappa \kappa' \right) \left( 1 - \lambda^2 - \mu^2 \right),  &2 P_{klrs} n_k n_l n'_r A_s &=  \mu \left( \kappa' \kappa^2 - 2 \lambda \kappa + \kappa' \right), \nonumber\\
 2 P_{klrs} n_k A_l n'_r n'_s &= \mu \left( \kappa \kappa'^2 - 2 \lambda \kappa' + \kappa \right), 
 &2 P_{klrs} n_k n_l n'_r n'_s &= \nu^2 \nu'^2 - 2 \mu^2, \label{eqn:Pcontralpha}\end{align}

\noindent and for $\sigma$,

\begin{align} 2 P_{klrs} n_k n'_l n'_r n_s &= \nu^2 \nu'^2,
 &2 P_{klrs} n_k n_l n'_r n_s &= \nu^2 \left( \lambda - \kappa \kappa' \right), \nonumber\\
 2 P_{klrs} n_k n_l n'_r n'_s &= \nu^2 \nu'^2 - 2 \mu^2,
&2 P_{klrs} n_k n'_l n'_r n'_s &= \nu'^2 \left( \lambda - \kappa \kappa' \right). \label{eqn:Pcontrsigma}\end{align}

\noindent Plugging these back into Eqs. (\ref{eqn:aP}) and (\ref{eqn:sP}) and simplifying, we find

\be \alpha(\Theta) = \frac{1}{4 \sin^4(\Theta)} \int d^2\Omega_{\mathbf{p}} \left[ (\lambda-\kappa \kappa')(1-\lambda^2) -\mu^2(1+\lambda) + \frac{2 \mu^2 (\lambda+\kappa)(\lambda+\kappa')}{(1+\kappa)(1+\kappa')} \right] = -\sigma(\Theta). \nonumber\ee

\noindent Noticing that we can do the integrals $\int d^2\Omega_{\mathbf{p}} \mu^2 = \frac{4\pi}{3} \sin^2\Theta$ and $\int d^2\Omega_{\mathbf{p}} \kappa \kappa' = \frac{4\pi}{3} \cos\Theta$, but that the last term is more complicated, we find

\be \alpha(\Theta) = -\sigma(\Theta) = \frac{\pi}{3} \frac{(\cos\Theta - 1)}{\sin^2\Theta} + \frac{1}{2 \sin^4\Theta} \int d^2\Omega_{\mathbf{p}} \frac{\mu^2 (\lambda+\kappa)(\lambda+\kappa')}{(1+\kappa)(1+\kappa')}. \label{eqn:alphaf}\ee

We can reduce the two-dimensional integral (\ref{eqn:alphaf}) to a one-dimensional integral by parametrizing $\mathbf{p}$ in spherical polar coordinates $\theta_p$ and $\phi_p$, choosing $\mathbf{n} = (0, \,\sin(\Theta/2), \,\cos(\Theta/2))$ and $\mathbf{n}' = (0, \,-\sin(\Theta/2), \,\cos(\Theta/2))$, and integrating over $\phi_p$. This gives

\be \alpha(\Theta) = -\sigma(\Theta) = \frac{\pi}{3} \frac{(\cos(\Theta) - 1)}{\sin^2(\Theta)} + \frac{\pi}{2 \sin^2\Theta} \int_0^{\pi} d\theta_p \sin\theta_p \left\{ \sin^2\theta_p + 8 \cos(\Theta/2) \left[ \cos\theta_p + \cos(\Theta/2) \right] \left[ \text{g}(\theta_p,\Theta) - 1 \right] \right\}, \label{eqn:alph1int}\ee

\noindent where

\be \text{g}(\theta_p,\Theta) = \frac{\left| \cos\theta_p + \cos(\Theta/2) \right|}{\left[ 1 + \cos\theta_p \cos(\Theta/2) \right]}. \ee

\noindent We perform the integral over $\theta_p$, and find the final form of the function $\alpha(\Theta)$

\be \alpha(\Theta) = -\sigma(\Theta) = \frac{\pi}{3 \sin^2\Theta} \left( 7 \cos\Theta - 5 \right) - \frac{32 \pi}{\sin^4\Theta} \ln\left( \sin(\Theta/2) \right) \sin^6(\Theta/2).\label{eqn:alphafin}\ee

\begin{figure}[t!]
\centering
\includegraphics[width=0.6\textwidth]{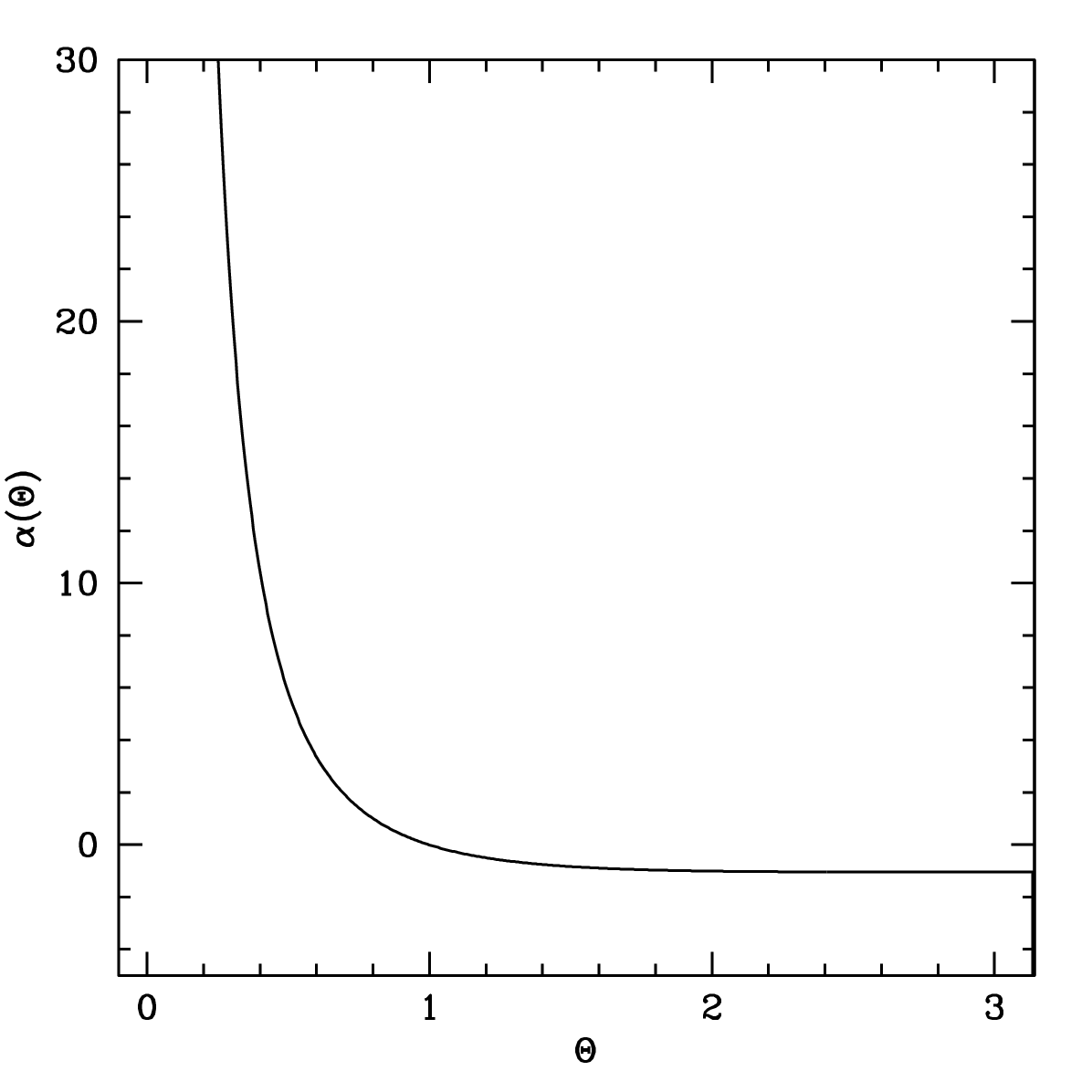}
\caption{Here we plot the function $\alpha(\Theta)$, the coefficient of $H_{ij}(\mathbf{n},\mathbf{n}')$ as shown in Eq. (\ref{eqn:Hdecompalph}),  as a function of the angle $\Theta$ between $\mathbf{n}$ and $\mathbf{n}'$.}
\label{fig:alpha}
\end{figure}

\noindent A plot of the function $\alpha(\Theta)$ is shown in Fig. \ref{fig:alpha}.

To summarize, we have now completed the calculation of the angular deflection correlation function. The final answer is given by Eq. (\ref{eqn:2ptH}), with $H_{ij}(\mathbf{n},\mathbf{n}')$ given from Eqs. (\ref{eqn:decomp2}) and (\ref{eqn:alphafin}) as

\be H_{ij}(\mathbf{n},\mathbf{n}') = \alpha(\Theta) \left( A_i A_j - B_i C_j \right). \label{eqn:Hdecompalph}\ee

\noindent Here the vectors $\mathbf{A}$, $\mathbf{B}$ and $\mathbf{C}$ are defined by Eqs. (\ref{eqn:ABC}), and $\alpha(\Theta)$ is given
  by Eq. (\ref{eqn:alphafin}).

\subsection{ Special Case: Coincidence }

As a check of our calculation, we can solve for the two-point correlation function exactly in the case that $\mathbf{n}=\mathbf{n}'$. Using Eqs. (\ref{eqn:R}), (\ref{eqn:P}) and (\ref{eqn:HR}), the integral simplifies to

\be H_{ij}(\mathbf{n},\mathbf{n}) = \frac{1}{4} \int d^2\Omega_{\mathbf{p}} \left[1 -  \left(\mathbf{p}\cdot\mathbf{n})^2\right)\right]\left( \delta_{ij} - n_i n_j \right).\nonumber\ee

\noindent We can solve this integral analytically, getting

\be H_{ij}(\mathbf{n},\mathbf{n}) = \frac{2 \pi}{3} \left( \delta_{ij} - n_i n_j \right). \label{eqn:Hcoinc}\ee

\noindent This corresponds to the limit of $\alpha(\Theta)(A^i A^j - B^i C^j)$ as $\mathbf{n}\rightarrow\mathbf{n}'$, with $\alpha(\Theta) = 2 \pi/(3 \Theta^2) + \mathcal{O}(\Theta^{-1})$ from Eq. (\ref{eqn:alphafin}). Inserting the coincidence limit (\ref{eqn:Hcoinc}) into the correlation function (\ref{eqn:2ptH}) yields the formula (\ref{eqn:2ptnn}) for the total rms angular fluctuations discussed in the introduction.

\section{ Spectrum of Angular Deflection Fluctuations \label{sec:spect}}

\subsection{ Overview }

In the previous section we computed the correlation function $\langle \delta n^i(\mathbf{n},t) \delta n^j(\mathbf{n}',t') \rangle$ as a function of the unit vectors $\mathbf{n}$ and $\mathbf{n}'$. However for many purposes it is more useful to perform a multipole decomposition of the angular deflection, and to compute the spectrum of fluctuations on different angular scales $l$, as is done with cosmic microwave background anisotropies. We decompose $\delta \mathbf{n}(\mathbf{n},t)$ as

\be \delta \mathbf{n}(\mathbf{n},t) = \sum_{l m} \delta n_{E lm}(t) \mathbf{Y}^E_{lm}(\mathbf{n}) + \delta n_{B lm}(t) \mathbf{Y}^B_{lm}(\mathbf{n}), \label{eqn:exp}\ee

\noindent where $\mathbf{Y}^E_{lm}$ and $\mathbf{Y}^B_{lm}$ are the electric- and magnetic-type transverse vector spherical harmonics defined by

\be \mathbf{Y}^E_{lm}(\mathbf{n}) = (l(l+1))^{-1/2} \mathbf{\nabla} Y_{lm}(\mathbf{n}),\quad \mathbf{Y}^B_{lm}(\mathbf{n}) = (l(l+1))^{-1/2} (\mathbf{n}\times\mathbf{\nabla}) Y_{lm}(\mathbf{n}). \label{eqn:Ydefn}\ee

\noindent We will show in this section that the statistical properties of the coefficients are given by

\be \langle \delta n_{Q lm}(t) \: \delta n_{Q' l'm'}(t')^* \rangle = \delta_{Q Q'} \delta_{l l'} \delta_{m m'} \int_0^{\infty} df \cos[2\pi f (t - t')] S_{Q l}(f) \label{eqn:2ptfnl} \ee

\noindent for $Q,Q' = E$ or $B$, for some spectrum $S_{Q l}(f)$, a function of frequency $f$ and of angular scale $l$. The formula (\ref{eqn:2ptfnl}) shows that different multipoles of the angular deflection are statistically independent, as required by spherical symmetry of the stochastic background. Also the electric-type and magnetic-type fluctuations are uncorrelated, as required by parity invariance of the stochastic background (see below).

The spectrum $S_{Q l}(f)$ is given by

\be S_{Q l}(f) = \frac{4\pi}{2l+1} \theta_{rms}^2 \frac{\sigma(f)}{f} g_Q \alpha_l^{QQ}. \label{eqn:SQlf} \ee

\noindent Here $\theta_{rms}^2$ is the total rms angular fluctuation squared, given by Eq. (\ref{eqn:2ptnn}) in the introduction. The function $\sigma(f)$ describes how the power is distributed in frequency. It is the same for all multipoles, is normalized according to $\int d(\, \ln f) \: \sigma(f) = 1$, and is given explicitly by Eq. (\ref{eqn:sigma}) in the introduction. The quantities $g_E$ and $g_B$ are the fraction of the total power in electric-type and magnetic-type fluctuations, and are $g_E = g_B = 1/2$. Finally the angular spectra $\alpha_l^{EE}$ and $\alpha_l^{BB}$ describe the dependence on angular scale, which is the same for all frequencies. They are normalized according to 

\be \sum_{l=2}^{\infty} \alpha_l^{QQ} = 1, \ee

\noindent and are the same for $E$- and $B$-modes, $\alpha_l^{EE} = \alpha_l^{BB}$. This spectrum is plotted in Fig \ref{fig:alphaQ} and the first 10 values are listed in Table \ref{tab:alphaQ}. We note that these coefficients are well fit by the power law $\alpha^{EE}_l = 32.34 \: l^{-4.921}$.

 Before proceeding with the derivation of the spectrum (\ref{eqn:2ptfnl}), we first derive from (\ref{eqn:2ptfnl}) the expression (\ref{eqn:thrms}) discussed in the introduction for the total fluctuation power. Squaring the expansion (\ref{eqn:exp}), taking an expected value, and then using (\ref{eqn:2ptfnl}) gives
 
 \begin{align} \langle \delta \mathbf{n}(\mathbf{n},t)^2 \rangle &= \sum_{Q l m} \sum_{Q' l' m'} \mathbf{Y}^Q_{lm}(\mathbf{n}) \mathbf{Y}^{Q'}_{l'm'}(\mathbf{n})^* \langle \delta n_{Q lm}(t) \delta n_{Q' l'm'}(t')^* \rangle \nonumber\\ &= \sum_{Ql} \int_0^{\infty} \frac{\sigma(f)}{f} \sum_{m=-l}^l \left| \mathbf{Y}^Q_{lm}(\mathbf{n}) \right|^2 \theta_{rms}^2 \: \frac{4 \pi}{2l+1} \: g_Q \; \alpha_l^{QQ}. \end{align}
 
 \noindent Using Uns$\ddot{\text{o}}$ld's theorem for vector spherical harmonics,
 
 \be \sum_{m=-l}^l \left| \mathbf{Y}^Q_{lm}(\mathbf{n}) \right|^2 = \frac{2l+1}{4\pi} ,\nonumber\ee
 
 \noindent gives
 
 \be \langle \delta \mathbf{n}(\mathbf{n},t)^2 \rangle = \sum_{Ql} \int_0^{\infty} \theta_{rms}^2 \: \frac{\sigma(f)}{f} \: g_Q \, \alpha_l^{QQ}, \ee
 
 \noindent which reduces to Eq. (\ref{eqn:thrms}). Note that using the normalization conventions for $\alpha_l^{QQ}$ and $\sigma(f)$ now gives $\langle \delta \mathbf{n}(\mathbf{n},t)^2 \rangle = \theta_{rms}^2 (g_E + g_B) = \theta_{rms}^2$, showing consistency of the definitions.
 
\subsection{ Derivation }

We now turn to a derivation of the spectrum (\ref{eqn:SQlf}). First we note that the vector spherical harmonics are transverse in the sense that $\mathbf{Y}^Q_{lm}(\mathbf{n})\cdot\mathbf{n} = 0$ for $Q=E,B$, and are orthogonal in the sense that

\be \int d^2\Omega_{\mathbf{n}} Y^Q_{lmi}(\mathbf{n}) Y^{Q'i*}_{l'm'}(\mathbf{n}) = \delta_{Q Q'} \delta_{l l'} \delta_{m m'}. \nonumber\ee

\noindent Using this orthogonality property, we can extract the coefficients of the expansion (\ref{eqn:exp})

\be \delta n_{Q lm}(t) = \int d^2\Omega_{\mathbf{n}} \delta n_i(\mathbf{n},t) Y^{Qi*}_{lm}(\mathbf{n}). \nonumber\ee

\noindent Thus we can write for the correlation function between two of these coefficients

\be \langle \delta n_{Qlm}(t) \delta n_{Q'l'm'}(t')^* \rangle = \int d^2\Omega_{\mathbf{n}} d^2\Omega_{\mathbf{n}'} Y^{Q*}_{lmi}(\mathbf{n}) Y^{Q'}_{l'm'j}(\mathbf{n}') \langle \delta n^i(\mathbf{n},t) \delta n^j(\mathbf{n}',t') \rangle, \label{eqn:twoptlm} \ee

\noindent or more explicitly, using Eq. (\ref{eqn:2ptH})

\be \langle \delta n_{Qlm}(t) \delta n_{Q'l'm'}(t')^* \rangle = \frac{3 H_0^2}{16 \pi^3} \int_0^{\infty} df \cos[2\pi f (t - t')] \: \frac{\Omega_{\rm gw}(f)}{f^3} \: C_{Q l m Q' l' m'}, \label{eqn:twopt} \ee

\noindent where

\be C_{Q l m Q' l' m'} = \int d^2\Omega_{\mathbf{n}} d^2\Omega_{\mathbf{n}'} Y^{Q*}_{lmi}(\mathbf{n}) Y^{Q'}_{l'm'j}(\mathbf{n}') H_{ij}(\mathbf{n},\mathbf{n}'). \label{eqn:C}\ee

We now argue that the EB cross-correlation vanishes. From Eq. (\ref{eqn:Ydefn}), we see that $\mathbf{Y}^E_{lm}(\mathbf{n})$ has the same parity under $\mathbf{n}\rightarrow-\mathbf{n}$ as $Y_{lm}(\mathbf{n})$, while the parity of $\mathbf{Y}^B_{lm}(\mathbf{n})$ is opposite. From Sec. \ref{sec:basis} above, $H_{ij}(\mathbf{n},\mathbf{n}')$ is invariant under both $\mathbf{n}\rightarrow-\mathbf{n}$ and $\mathbf{n}'\rightarrow-\mathbf{n}'$. Thus, if $Q=E$, $Q' = B$ in Eq. (\ref{eqn:twopt}), the integral will be symmetric under $\mathbf{n}\rightarrow-\mathbf{n}$ but antisymmetric under $\mathbf{n}'\rightarrow-\mathbf{n}'$, causing the integral over $d^2 \Omega_{\mathbf{n}'}$ to vanish. Therefore, $EB$ cross correlations vanish, and we need only calculate the $EE$ and $BB$ correlation functions.

\subsubsection{ EE correlation }

Inserting the definition (\ref{eqn:Ydefn}) of the electric vector spherical harmonics and the formula (\ref{eqn:Hdecompalph}) for $H_{ij}$ into Eq. (\ref{eqn:C}) and integrating by parts, we obtain

\be C_{E l m E' l' m'} = \frac{1}{l(l+1)} \int d^2\Omega_{\mathbf{n}} d^2\Omega_{\mathbf{n}'} Y^*_{lm}(\mathbf{n}) Y_{l'm'}(\mathbf{n}') \beta^{EE}(\Theta), \label{eqn:C2} \ee

\noindent where the function $\beta^{EE}$ is given by

\be \beta^{EE}(\Theta) = \nabla_i \nabla'_j \left[ H_{ij}(\mathbf{n},\mathbf{n}') \right] = \nabla_i \nabla'_j \left\{\alpha(\Theta) \left[ A_i A_j - B_i C_j \right]\right\}. \label{eqn:bEEdef1}\ee

\noindent Here $\nabla_i$ and $\nabla'_j$ denote normal three dimensional derivatives with respect to $\mathbf{x}$ and $\mathbf{x}'$, where $\mathbf{n} = \mathbf{x}/|\mathbf{x}|$ and $\mathbf{n}' = \mathbf{x}'/|\mathbf{x}'|$. Integration by parts on the unit sphere of this derivative operator is valid as long as the radial component of the integrand vanishes, from the identity $\nabla_i v^i = \partial_r v^r + 2 v_r/r + \nabla_A v^A$, where $\nabla_A$ denotes a covariant derivative on the unit sphere. It can be checked that the radial components do vanish in the above computation.

Next, we expand the function $\beta^{EE}$ in terms of Legendre polynomials, and use the spherical harmonic addition theorem, which gives

\begin{align} \beta^{EE}(\Theta) &= \sum_l \beta^{EE}_l P_l(\cos\Theta) \nonumber\\ &= \sum_{lm} \frac{4 \pi}{2l+1} \beta^{EE}_l Y_{lm}(\mathbf{n}) Y_{lm}(\mathbf{n}')^*\label{eqn:bEEexp}\end{align}

\noindent Inserting this into Eq. (\ref{eqn:C2}) and using the orthogonality of spherical harmonics gives

\be C_{E l m E' l' m'} = \delta_{l l'} \delta_{m m'} \frac{1}{l (l+1)} \frac{4 \pi}{2l+1} \beta^{EE}_l. \ee

\noindent Inserting this into Eq. (\ref{eqn:twopt}) now yields the correlation function given by Eqs. (\ref{eqn:2ptfnl}) and (\ref{eqn:SQlf}), and using the definitions (\ref{eqn:2ptnn}) and (\ref{eqn:sigma}) of $\theta_{rms}^2$ and $\sigma(f)$ allows us to read off the electric multipole spectrum

\be g_E \: \alpha_l^{EE} = \frac{3}{4 \pi l (l+1)} \beta_l^{EE}. \label{eqn:bEEalph}\ee

\noindent We will show below that $g_E = 1/2$.

It remains to explicitly evaluate the function $\beta^{EE}(\Theta)$ defined in Eq. (\ref{eqn:bEEdef1}) and evaluate its expansion coefficients. We have

\be \beta^{EE}(\Theta) \equiv \nabla_i \nabla'_j \left[\alpha(\Theta)T^{ij}\right] = \left[\nabla_i \nabla'_j \alpha(\Theta)\right] T^{ij} + \left[\nabla_i \alpha(\Theta)\right] \left(\nabla'_j T^{ij}\right) + \left[\nabla'_j \alpha(\Theta)\right] \left(\nabla_i T^{ij}\right) + \alpha(\Theta)\left(\nabla_i \nabla'_j T^{ij}\right), \label{eqn:bEEdef}\ee

\noindent where we have defined $T^{ij} = \left(A^i A^j(\mathbf{n},\mathbf{n}') - B^i C^j(\mathbf{n},\mathbf{n}')\right)$. Using $A^i = \epsilon^{ijk}n_j n'_k$, $B^i = (\mathbf{n}\cdot\mathbf{n'}) n^i - n'^i$, $C^i = (\mathbf{n}\cdot\mathbf{n'}) n'^i - n^i$, we can write the tensor $T^{ij}$ in Cartesian coordinates as 

\be T^{ij} = \epsilon^{ikl} \epsilon^{jrs} n_k n'_l n_r n'_s - \left( (\mathbf{n}\cdot\mathbf{n}') n^i - n'^i \right) \left( (\mathbf{n}\cdot\mathbf{n}') n'^j - n^j \right). \nonumber\ee

\noindent Using $\nabla_i n_j = \delta_{ij} - n_i n_j$, $\nabla'_i n'_j = \delta_{ij} - n'_i n'_j $, $\nabla'_i n^j = \nabla_i n'^j  = 0$, and $\nabla_l \epsilon^{ijk} = \nabla'_l \epsilon^{ijk} = 0$, we calculate the derivatives

\begin{align} \nabla_i T^{ij} = \left(1 - 3 (\mathbf{n}\cdot\mathbf{n}')\right) \left( (\mathbf{n}\cdot\mathbf{n}') n'^j - n^j \right),& \quad\quad \nabla'_j T^{ij} = \left(1 - 3 (\mathbf{n}\cdot\mathbf{n}')\right) \left( (\mathbf{n}\cdot\mathbf{n}') n^i - n'^i \right), \nonumber\\ \nabla_i \nabla'_j T^{ij} =&-9 (\mathbf{n}\cdot\mathbf{n}')^2 + 2 (\mathbf{n}\cdot\mathbf{n}') + 3 . \label{eqn:dT}\end{align}

For the gradients of $\alpha$, we use the fact that $\cos(\Theta) = \mathbf{n}\cdot\mathbf{n'}$, so that $- \sin(\Theta) \nabla_i \Theta = n'_i - (\mathbf{n}\cdot\mathbf{n}') n_i$, and similarly for $\nabla'_j$. Thus, we find

\begin{align} \nabla_i \alpha(\Theta)&= - \alpha'(\Theta) \frac{n'_i - (\mathbf{n}\cdot\mathbf{n}') n_i}{\sin(\Theta)}, \quad\quad \nabla'_j \alpha(\Theta)= - \alpha'(\Theta) \frac{n_j - (\mathbf{n}\cdot\mathbf{n}') n'_j}{\sin(\Theta)}\nonumber\\
\nabla_i \nabla'_j \alpha(\Theta)&= \alpha'(\Theta) \left\{ \frac{\delta_{ij} - n_i n_j - n'_i n'_j + (\mathbf{n}\cdot\mathbf{n}')n_i n'_j}{-\sin(\Theta)} + \frac{\cos(\Theta)\left[n'_i - (\mathbf{n}\cdot\mathbf{n}') n_i\right]\left[ n_j - (\mathbf{n}\cdot\mathbf{n}') n'_j \right]}{-\sin^3(\Theta)} \right\}  \nonumber\\
&+ \alpha''(\Theta) \frac{\left[n'_i - (\mathbf{n}\cdot\mathbf{n}') n_i\right]\left[ n_j - (\mathbf{n}\cdot\mathbf{n}') n'_j \right]}{\sin^2(\Theta)}. \label{eqn:dalph} \end{align}

\noindent Plugging Eqs. (\ref{eqn:dT}) and (\ref{eqn:dalph}) into Eq. (\ref{eqn:bEEdef}), we get

\begin{align} \beta^{EE}(\Theta) =& \left[ -9 \cos^2(\Theta) + 2 \cos(\Theta) + 3 \right] \alpha(\Theta) - \sin^2(\Theta)\alpha''(\Theta) \nonumber \\& + \left[ 1 - 6 \cos(\Theta)\right] \sin(\Theta)\alpha'(\Theta). \label{eqn:brack2}\end{align}

\noindent Next, we insert the expression (\ref{eqn:alphafin}) for $\alpha(\Theta)$ to obtain

\be \beta^{EE}(\Theta) = \frac{4 \pi}{3} \Big( 4 + \left( 1 - \cos\Theta \right) \left\{ 12 \ln\left[ \sin(\Theta/2) \right] - 1 \right\} \Big). \label{eqn:beefin}\ee

\noindent We numerically compute the coefficients $\beta^{EE}_l$ of the Legendre polynomial expansion (\ref{eqn:bEEexp}) of $\beta^{EE}(\Theta)$, and from them compute $\alpha_l^{EE}$ using Eq. (\ref{eqn:bEEalph}). The result is plotted in Fig. \ref{fig:alphaQ} and tabulated in Table \ref{tab:alphaQ}.

\subsubsection{ BB correlation }

We now calculate the $BB$ correlation in a similar manner to the $EE$ case above. Inserting into Eq. (\ref{eqn:C}) the definition (\ref{eqn:Ydefn}) of magnetic vector spherical harmonics and integrating by parts, we find

\be C_{B l m B l' m'} = \frac{1}{l (l+1)} \int d^2\Omega_{\mathbf{n}} d^2\Omega_{\mathbf{n}'} Y^*_{lm}(\mathbf{n}) Y_{l'm'}(\mathbf{n}') \beta^{BB}(\Theta), \nonumber\ee

\noindent where

\be \beta^{BB}(\Theta) = \nabla_l \nabla'_p \left[\epsilon_{ikl}\epsilon_{jmp} n_k n'_m \alpha(\Theta) T_{ij} \right]. \label{eqn:bBB}\ee

\noindent As before, we can derive from here the form (\ref{eqn:2ptfnl}) and (\ref{eqn:SQlf}) of the spectrum, with $\alpha_l^{BB}$ given by

\be g_B \: \alpha_l^{BB} = \frac{3}{4 \pi l (l+1)} \beta_l^{BB}. \nonumber\ee

We now show that $\beta^{BB}(\Theta) = \beta^{EE}(\Theta)$, from which it follows that $g_E = g_B = 1/2$ and that $\alpha_l^{EE} = \alpha_l^{BB}$. To see this we evaluate the cross products in (\ref{eqn:bBB}) using $\mathbf{n}\times\mathbf{A} = \mathbf{B}$, $\mathbf{n}\times\mathbf{B} = -\mathbf{A}$, $\mathbf{n}'\times\mathbf{C} = \mathbf{A}$. This gives

\be \epsilon_{ikl} \epsilon_{jmp} n_k n'_m H_{ij} = H_{lp}, \nonumber\ee

\noindent and using the definitions (\ref{eqn:bEEdef1}) and (\ref{eqn:bBB}) of $\beta^{EE}$ and $\beta^{BB}$, it follows that $\beta^{BB} = \beta^{EE}$.

\acknowledgments
LGB acknowledges the support of the NSF Graduate Fellowship Program. EF thanks the Theoretical Astrophysics Including Relativity Group at Caltech and the Department of Applied Mathematics and Theoretical Physics at the University of Cambridge for their hospitality as this paper was being written. This research was supported in part by NSF Grants No. PHY-0757735 and PHY-0555216.

%
%

\bibliography{ref}
%
%
%
%
%
%
%
\appendix
\section{NonLocal Dependence of Deflection Angle on Gravitational Waves}

It is sometimes claimed in the literature that the deflection angle, in the distance source limit, depends only on the GWs in the vicinity of the source and the observer. A similar claim is often made for the frequency perturbation caused by GWs which is targeted in pulsar timing searches. Strictly speaking,  these claims are not true: it is possible to have a nonzero deflection when the GW field vanishes in a neighborhood of the source and of the observer, even in the distant source limit. However, this type of circumstance requires a considerable fine tuning, so the claims are colloquially valid.

To see this, it is sufficient to consider the simple model of a scalar field $h(t,x)$ in $1+1$ dimensions, obeying  $\left(\partial_t^2-\partial_x^2\right) h = 0$. A functional of $h(t,x)$ that is qualitatively similar to the deflection angle formula (\ref{eqn:2ptH}) is

\be \Delta \theta = c_R h_R(t-x_{\rm obs}) + c_L h_L(t +x_{\rm obs}). \label{eqn:appndx}\ee

\noindent where we have decomposed the field into left-moving and right-moving pieces,

\be h(t,x) = h_R(t-x) + h_L(t+x),\ee

\noindent and $c_R$ and $c_L$ are fixed coefficients with $c_L \neq c_R$. In each sector (right-moving and left-moving), the quantity (\ref{eqn:appndx}) depends only on the field evaluated at the location of the observer $x=x_{obs}$. However, the sum does not. If we specify the field in terms of its initial data $h(t,x)$ and $\dot{h}(t,x)$ at time $t$, t, and specialize to initial data of compact support, we get

\be \Delta \theta(t) = \frac{1}{2} (c_L + c_R) h(t,x_{obs}) + \frac{1}{2} (c_L - c_R) \int_{-\infty}^{x_{obs}} dx \dot{h}(t,x). \ee

\noindent It is clearly possible to choose $h$ and $\dot{h}$ to vanish in a neighborhood of $x=x_{obs}$ and still have $\Delta \theta \neq 0$, for $c_L \neq c_R$.

\end{document}